\documentclass{emulateapj}

\shorttitle{Mid-IR Properties of Starburst Galaxies}
\shortauthors{Brandl et al.}
\newcommand{\2}{H{\sc ii}\ }

\begin{document}

\title{The Mid-IR Properties of Starburst Galaxies
       from {\em Spitzer-IRS} Spectroscopy}


%
%
\author{
B.R. Brandl$^1$,
J. Bernard-Salas$^2$,
H.W.W. Spoon$^2$\altaffilmark{\dag},
D. Devost$^2$,
G.C. Sloan$^2$,
S. Guilles$^2$,
Y. Wu$^2$,
J.R. Houck$^2$,
L. Armus$^3$,
D.W. Weedman$^2$,
V. Charmandaris$^{4,2,}$\altaffilmark{\ddag},
P.N. Appleton$^3$,
B.T. Soifer$^3$,
L. Hao$^2$,
J.A. Marshall$^2$,
S.J. Higdon$^2$,
\& T.L. Herter$^2$
}
\affil{$^1$Leiden University, P.O. Box 9513, 2300 RA Leiden, The Netherlands}
\email{brandl@strw.leidenuniv.nl}
\affil{$^2$Cornell University, Astronomy Department, Space Sciences Building,
       Ithaca, NY 14853}
\affil{$^3$Caltech Spitzer Science Center, MS 314-6, Pasadena, CA 91125}
\affil{$^4$University of Crete, Department of Physics, P.O. Box 2208
       GR-71003, Heraklion, GREECE}
\altaffiltext{\dag}{Spitzer Fellow}
\altaffiltext{\ddag}{Chercheur Associ\'e, Observatoire de Paris, F-75014,
        Paris, France}

\begin{abstract}
We present $5 - 38\mu$m mid-infrared spectra at a spectral resolution
of $R\approx 65 - 130$ of a large sample of 22 starburst nuclei taken
with the Infrared Spectrograph {\sl IRS} on board the Spitzer Space
Telescope.
The spectra show a vast range in starburst SEDs.  The silicate
absorption ranges from essentially no absorption to heavily obscured
systems with an optical depth of $\tau_{9.8\mu m}\sim 5$.  The
spectral slopes can be used to discriminate between starburst and AGN
powered sources.  The monochromatic continuum fluxes at $15\mu$m and
$30\mu$m enable a remarkably accurate estimate of the total infrared
luminosity of the starburst.
We find that the PAH equivalent width is independent of the total
starburst luminosity $L_{IR}$ as both continuum and PAH feature scale
proportionally.  However, the luminosity of the $6.2\mu$m feature
scales with $L_{IR}$ and can be used to approximate the total infrared
luminosity of the starburst.
Although our starburst sample covers about a factor of ten difference
in the [Ne\,III]\,/\,[Ne\,II] ratio, we found no systematic
correlation between the radiation field hardness and the PAH
equivalent width or the $7.7\mu$m\,/\,$11.3\mu$m PAH ratio.
These results are based on spatially integrated diagnostics over an
entire starburst region, and local variations may be ``averaged out''.
It is presumably due to this effect that unresolved starburst nuclei
with significantly different global properties appear spectrally as
rather similar members of one class of objects.

\end{abstract}

\keywords{Telescopes: {\em Spitzer}, 
          Galaxies: ISM, starburst, star clusters
	  ISM: dust, extinction, HII regions}


\section{Introduction}
\label{secintro}

Many nearby galaxies show dramatically increased rates of star
formation compared with the Milky Way. In such "Starburst" galaxies
(e.g., \citet{wee81}), the primary energy source is driven by high
nuclear star formation rates, rapid nuclear gas depletion time-scales
and high supernovae rates. Such starburst systems often, though not
exclusively, occur in interacting and colliding systems. Since
collisions and interactions are believed to be a fundamental part of
the evolution of galaxies by hierarchical growth, the full
characterization of starburst galaxies is of the great importance in
measuring and quantifying the global history of star formation over
cosmic time.  Determining the average mid-infrared (mid-IR) spectral
properties, and the range of observed behavior within the starburst
class at low redshift is vital for interpreting spectra of higher
redshift IR sources, providing complementary spectral templates to
those parallel Spitzer studies of the more extreme ULIRG systems
(e.g., \citet{arm06b}) and AGN (e.g., \citet{wee05}).

The term ``starburst galaxy'' is commonly used to describe an
apparently well-defined class of objects, although starbursts can be
found in the most diverse conditions, ranging from low-pressure dwarf
galaxies to high-pressure nuclear starbursts.  Their observed
properties are expected to depend on numerous parameters such as the
initial stellar mass function (IMF), the duration and epoch of the
individual starburst(s), the metallicity of the ISM, the size and
distribution of the dust grains, the strength of the magnetic fields,
gas pressure and temperature of the ISM, galactic shear, total
luminosity, and total mass.  Furthermore, nearby starbursts, for which
high resolution imaging is possible, have revealed complex
substructures -- in both stellar distributions and ISM -- ranging from
ultra-compact H\,{\sc ii} regions (UCHIIR) to large complexes of super
star clusters (SSC), suggesting small-scale variations of the
observables across a starburst region.

We use the low resolution mode of the Infrared
Spectrograph\footnote{The {\sl IRS} was a collaborative venture
between Cornell University and Ball Aerospace Corporation funded by
NASA through the Jet Propulsion Laboratory and the Ames Research
Center.} ({\sl IRS}) \citep{hou04} on board the {\sl Spitzer Space
Telescope} \citep{wer04} to observe the central regions of 22
starburst galaxies.  Our objects represent a sample of ``classical''
starbursts for which a wealth of literature exists.  The sample
includes both purely starburst and starbursts with weak AGN activity
(as determined from X-ray, optical, or radio observations).  The
summary in Table~\ref{tabgenprop} lists the observed targets, their
general properties, the classifications we adopt, and the references
from which they are derived.  The continuous $5-38\mu$m {\sl IRS}
spectra include the silicate bands around $10\mu$m and $18\mu$m, a
large number of PAH emission features, and information on the slope of
the spectral continuum.

\begin{deluxetable*}{l r r l r r c r r r r}
\centering
\tabletypesize{\footnotesize} 
\tablecaption{General Properties} 
\tablewidth{0pt} 
\tablehead{ 
\colhead{Name} & 
\colhead{$\alpha_{\mbox{J2000}}$\tablenotemark{a}} &
\colhead{$\delta_{\mbox{J2000}}$\tablenotemark{a}} & 
\colhead{Type} & 
\colhead{Refs.\tablenotemark{b}} &
\colhead{D\tablenotemark{c}} & 
\colhead{$\log(L_{IR})$\tablenotemark{d}} &
\colhead{$S_{12}$\tablenotemark{e}} & 
\colhead{$S_{25}$\tablenotemark{e}} & 
\colhead{$S_{60}$\tablenotemark{e}} & 
\colhead{$S_{100}$\tablenotemark{e}}\\
& & & & & \colhead{[Mpc]} & 
\colhead{[$L_\odot$]} &
\colhead{[Jy]} & 
\colhead{[Jy]} & 
\colhead{[Jy]} & 
\colhead{[Jy]}
}
\startdata 
IC   342 &  3 46 48.51  &   +68 05 46.0 & SB & 13,24,25   &  4.6 & 10.17 & 14.92 & 34.48 & 180.80 & 391.66 \\
Mrk   52 & 12 25 42.67  &   +00 34 20.4 & SB & 2,9,17     & 30.1 & 10.14 & 0.28 &  1.05 &  4.73 &   5.68 \\
Mrk  266 & 13 38 17.69  &   +48 16 33.9 & SB+Sy2 & 14,20  & 115.8 & 11.49 & 0.32 &  1.07 &  7.25 &  10.11 \\
NGC  520 &  1 24 35.07  &   +03 47 32.7 & SB & 4,10,12,13 & 30.2 & 10.91 & 0.90 &  3.22 & 31.52 &  47.37 \\
NGC  660 &  1 43 02.35  &   +13 38 44.4 & SB+LINER & 10     & 12.3 & 10.49 & 3.05 &  7.30 & 65.52 & 114.74 \\
NGC 1097 &  2 46 19.08  & $-$30 16 28.0 & SB+Sy1 & 19,23  & 16.8 & 10.71 & 2.96 &  7.30 & 53.35 & 104.79 \\
NGC 1222 &  3 08 56.74  & $-$02 57 18.5 & SB & 1,2        & 32.3 & 10.60 & 0.50 &  2.28 & 13.06 &  15.41 \\
NGC 1365 &  3 33 36.37  & $-$36 08 25.5 & SB+Sy2 & 19,26  & 17.9 & 11.00 & 5.12 & 14.28 & 94.31 & 165.67 \\
NGC 1614 &  4 33 59.85  & $-$08 34 44.0 & SB & 29,30      & 62.6 & 11.60 & 1.38 & 7.50  & 32.12 & 34.32  \\
NGC 2146 &  6 18 37.71  &   +78 21 25.3 & SB & 7,10       & 16.5 & 11.07 & 6.83 & 18.81 & 146.69 & 194.05 \\
NGC 2623 &  8 38 24.08  &   +25 45 16.9 & SB & 13,22      & 77.4 & 11.54 & 0.21 &  1.81 & 23.74 &  25.88 \\
NGC 3256 & 10 27 51.27  & $-$43 54 13.8 & SB & 9,24,25    & 35.4 & 11.56 & 3.57 & 15.69 & 102.63 & 114.31 \\
NGC 3310 & 10 38 45.96  &   +53 30 05.3 & SB & 9,10       & 19.8 & 10.61 & 1.54 &  5.32 & 34.56 &  44.19 \\
NGC 3556 & 11 11 30.97  &   +55 40 26.8 & SB & 7,10       & 13.9 & 10.37 & 2.29 &  4.19 & 32.55 &  76.90 \\
NGC 3628 & 11 20 17.02  &   +13 35 22.2 & SB+LINER & 10,21  & 10.0 & 10.25 & 3.13 &  4.85 & 54.80 & 105.76 \\
NGC 4088 & 12 05 34.19  &   +50 32 20.5 & SB & 3,5,10     & 13.4 & 10.25 & 2.06 &  3.45 & 26.77 &  61.68 \\
NGC 4194 & 12 14 09.64  &   +54 31 34.6 & SB & 2,9,17     & 40.3 & 11.06 & 0.99 &  4.51 & 23.20 &  25.16 \\
NGC 4676 & 12 46 10.10  &   +30 43 55.0 & SB & 15,16      & 94.0 & 10.88 & 0.11 &  0.33 &  2.67 &   5.18 \\
NGC 4818 & 12 56 48.90  & $-$08 31 31.1 & SB & 5,17       &  9.4 & 09.75 & 0.96 &  4.40 & 20.12 &  26.60 \\
NGC 4945 & 13 05 27.48  & $-$49 28 05.6 & SB+Sy2 & 7,11,31  & 3.9 & 10.48 & 27.74 & 42.34 & 625.46 & 1329.70 \\
NGC 7252 & 22 20 44.77  & $-$24 40 41.8 & SB & 6,15,18    & 66.4 & 10.75 & 0.24 &  0.43 &  3.98 &   7.02 \\
NGC 7714 & 23 36 14.10  &   +02 09 18.6 & SB & 8,27,28    & 38.2 & 10.72 & 0.47 &  2.88 & 11.16 &  12.26\\  
\enddata
\label{tabgenprop}
\tablecomments{Mrk 52 = NGC 4385, Mrk 266 = NGC 5256}
\tablenotetext{a}{Commanded coordinates of the slit center.
               $\alpha$ is given in (h~m~s), $\delta$ is given in (d~m~s).}
\tablenotetext{b}{References:
                  1--\citet{ash95}, 2--\citet{bal83},
                  3--\citet{ben04}, 4--\citet{bes03},
                  5--\citet{dev89}, 6--\citet{fri94},
                  7--\citet{gao04}, 8--\citet{gon99},
                  9--\citet{hec98}, 10--\citet{ho97},
                  11--\citet{iwa93}, 12--\citet{jos85},
                  13--\citet{kee84}, 14--\citet{lev01},
                  15--\citet{liu95}, 16--\citet{lon84},
                  17--\citet{may04}, 18--\citet{mil97},
                  19--\citet{osm74}, 20--\citet{ost83},
                  21--\citet{rob04}, 22--\citet{smi98},
                  23--\citet{sto03}, 24--\citet{tho00},
                  25--\citet{ver03}, 26--\citet{ver80},
                  27--\citet{wee81}, 28--\citet{bra04},
                  29--\citet{alo01}, 30--\citet{ket92},
                  31--\citet{spo00}}
\tablenotetext{c}{Distances adopted from \citet{san03}, except for Mrk~52, 
                  NGC~4676 and NGC~7252, which were derived from measured 
                  redshifts via
                  $D = \frac{cz}{H_{\mbox{0}}}(1+\frac{z}{2})$, 
                  assuming $H_{\mbox{0}} = 71$\,km\,s$^{-1}$ Mpc$^{-1}$.}
\tablenotetext{d}{Total $8 - 1000\mu$m infrared luminosity of the entire 
                  galaxy, adopted from \citet{san03}, except for Mrk~52, 
                  NGC~4676 and NGC~7252, which were derived from measured 
                  IRAS fluxes via
                  $L_{IR} = 312700\cdot D^2\cdot 1.8\cdot (13.48S_{12\mu m} + 
                  5.16S_{25\mu m} + 2.58S_{60\mu m} + S_{100\mu m})$ where
                  $S_{\lambda}$ is in Jy.} 
\tablenotetext{e}{{\sl IRAS} flux densities at 12, 25, 60 and $100\mu$m 
                  of the entire galaxy, adopted from \citet{san03}.}
\end{deluxetable*}

Numerous mid-IR studies of starbursts have been conducted with
{\sl ISO-SWS}, {\sl ISOCAM}, or {\sl ISOPHOT}, see for instance, 
\citet{rig96},
\citet{lut98},
\citet{rig99},
\citet{dal00},
\citet{hel00},
\citet{lau00},
\citet{stu00}, 
\citet{tho00}, 
\citet{cha01},
\citet{for03},
\citet{lu03},
\citet{ver03},
\citet{tac05}, and
\citet{mad06}.
An overview of many {\sl ISO} results is given in \citet{gen00}.  While
the {\sl ISO} observations provided great new insights in the spectral
characteristics of individual starbursts, the sample of continuum
spectra remained rather small or was limited to shorter wavelengths or
narrow bandwidth scans of the strongest emission lines.

In this paper we will address the question of the mid-IR homogenity of
the classical starburst class\footnote{In this paper the terms
`starburst', `starburst galaxy', and `starburst nuclei' all refer to
the central, sub-kiloparsec regions of galaxies with significantly
enhanced starburst activity}, and attempt to investigate how specific
spectral features (especially the mid-IR PAH bands) vary with the
total UV continuum flux, UV hardness ratio and dust extinction within
the starburst nucleus.  We will also investigate how the shape of the
continuum depends on the luminosity source, and if the total
luminosity affects the observed spectral shapes, i.e., to what extent
starbursts can be scaled up.  We investigate the role of dust, and how
well PAH emission correlates with the rate of star formation.

The outline of the paper is as follows: First we give a detailed
description of the observations and the data reduction and
calibration.  In Section~\ref{secanaly} we discuss how the relevant
spectral features (SED, PAHs, silicate features) have been measured.
The main focus will be on the discussion of the numerous results in
section~\ref{secdiscus}, followed by a summary.
We note that our sample has also been observed with the {\sl IRS}
high resolution modules, revealing the large, comprehensive zoo of
strong and faint fine-structure lines in the $10-38\mu$m wavelength
range. This analysis is complementary to our above science goals and
will be presented in a subsequent paper by \citet{dev06}.


\section{Observations and Data Reduction}

\subsection{Observations}

\begin{figure*}[tb]
\plotone{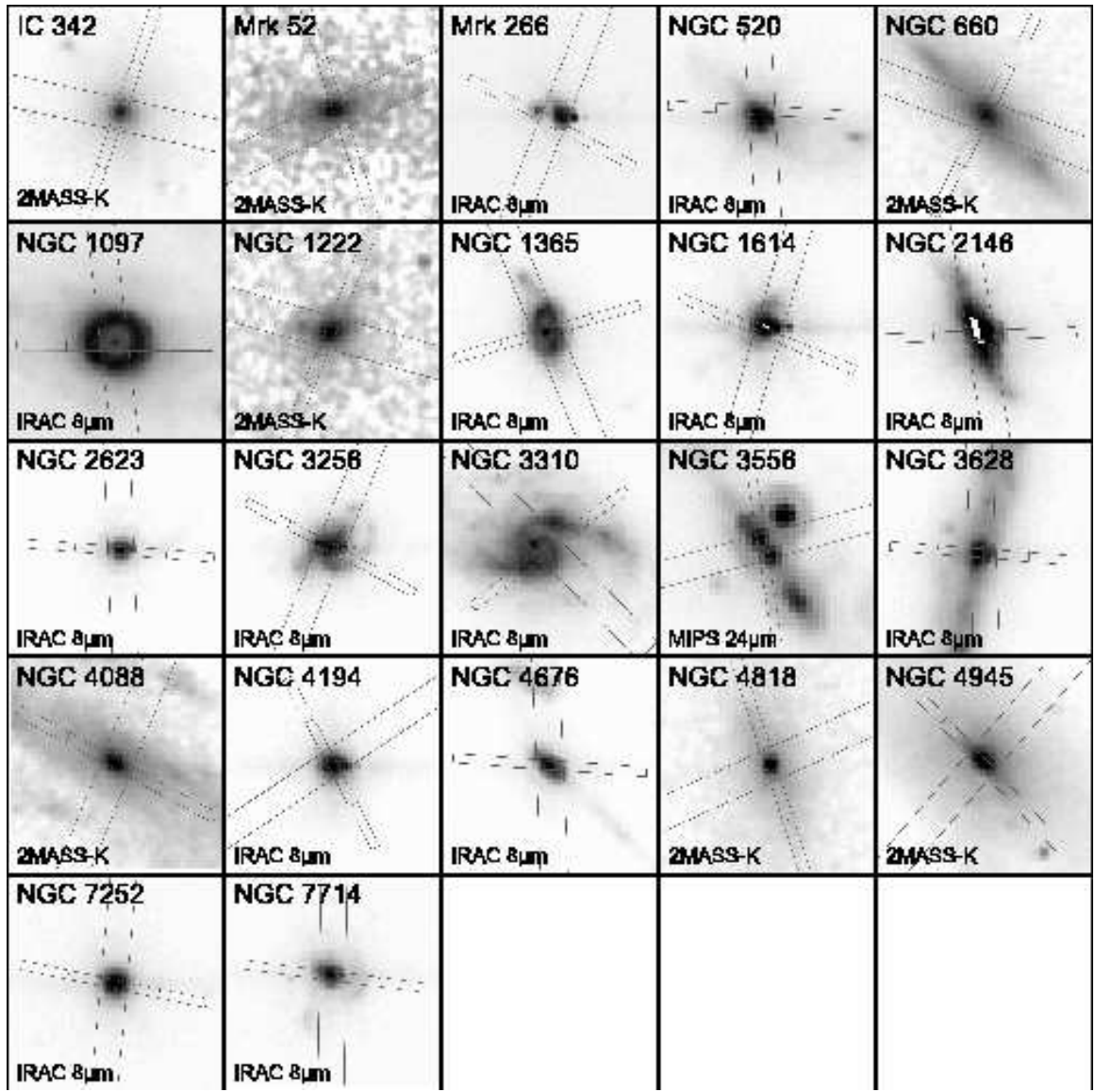}
\caption{The positions of the SL and LL slits for the given observing
         date overplotted on IRAC $8\mu$m, MIPS $24\mu$m or 2MASS
         K-band images for all galaxies in our sample. The scale
         is logarithmic and overemphasizes the real flux
         distribution. For discussion see section~\ref{secstitching}.
         \label{figslitoverlays}}
\end{figure*}

We observed all targets with the two low-resolution modules ($R\approx
65 - 130$) of the {\sl IRS}.  The slit widths are about $3.6''$
from $5 - 15\mu$m and $10.5''$ from $15 - 38\mu$m.  The observations
were made within the first year of the {\sl Spitzer Space Telescope}
mission (see Table~\ref{tabobsprop}) as part of the {\sl IRS}
guaranteed time observing program.  The data were taken using standard
{\sl IRS} ``Staring Mode'' Astronomical Observing Templates (AOT).  In
most cases, a high accuracy {\sl IRS} blue peak-up, offsetting from a
nearby 2MASS star, was performed to achieve the intended pointing
accuracy.  The central coordinates of the slits for these observations
were derived from 2MASS images.  Table~\ref{tabobsprop} lists the
observing parameters for all targets. Fig.~\ref{figslitoverlays} shows
the slit positions relative to the galaxies.

\begin{deluxetable*}{l c r c r r r r r r r}
\centering 
\tabletypesize{\footnotesize}
\tablecaption{Observational Parameters} 
\tablewidth{0pt} 
\tablehead{ 
\colhead{Name} &
\colhead{AOR-key} &
\colhead{Obs.date} & 
\colhead{$t_{SL}$\tablenotemark{a}} & 
\colhead{$t_{LL}$\tablenotemark{a}} & 
\multicolumn{5}{c}{Stitching factors\tablenotemark{b}} & 
\colhead{FF\tablenotemark{c}}
}
\startdata 
IC   342 & 9072128 & 2004-03-01 & $2\times 14 $ & $4\times 6 $  & 2.22 & 1.62 & 1.89 & 0.95 & 0.95 & 0.47 \\
Mrk   52 & 3753216 & 2004-01-08 & $4\times 6 $  & $4\times 6 $  & 1.43 & 1.47 & 1.57 & 0.95 & 0.95 & 0.84 \\
Mrk  266 & 3755264 & 2004-01-08 & $2\times 14 $ & $2\times 14 $ & 1.89 & 1.89 & 1.79 & 0.94 & 1.00 & 0.58 \\
NGC  520 & 9073408 & 2004-07-13 & $3\times 14 $ & $3\times 14 $ & 5.26 & 3.53 & 4.11 & 0.95 & 1.00 & 0.73 \\
NGC  660 & 9070848 & 2004-08-07 & $4\times 6 $  & $4\times 6 $  & 1.52 & 1.33 & 1.45 & 0.95 & 1.00  & 0.73 \\
NGC 1097 & 3758080 & 2004-01-08 & $2\times 14 $ & $4\times 6 $  & 0.69 & 0.72 & 0.75 & 0.69 & 0.79  & 0.21 \\
NGC 1222 & 9071872 & 2004-08-10 & $2\times 14 $ & $2\times 14 $ & 1.52 & 1.18 & 1.38 & 0.93 & 0.95  & 0.76 \\
NGC 1365 & 8767232 & 2004-01-04 & $4\times 6 $  & $4\times 6 $  & 1.56 & 1.09 & 1.34 & 0.88 & 0.94 & 0.37 \\
NGC 1614 & 3757056 & 2004-02-06 & $2\times 14 $ & $4\times 6 $  & 1.27 & 1.02 & 1.19 & 0.88 & 0.88 & 0.52 \\
NGC 2146 & 9074432 & 2004-02-28 & $4\times 6 $  & $4\times 6 $  & 2.50 & 2.00 & 2.28 & 0.85 & 0.90 & 0.55 \\
NGC 2623 & 9072896 & 2004-04-19 & $2\times 14 $ & $2\times 14 $ & 1.25 & 1.12 & 1.25 & 1.05 & 1.05 & 0.94 \\
NGC 3256 & 9073920 & 2004-05-13 & $4\times 6 $  & $4\times 6 $  & 1.57 & 1.20 & 1.42 & 1.00 & 1.00 & 0.75 \\
NGC 3310 & 9071616 & 2004-04-19 & $2\times 14 $ & $4\times 6 $  & 0.69 & 0.76 & 0.76 & 0.84 & 0.90 & 0.28 \\
NGC 3556 & 9070592 & 2004-04-18 & $2\times 14 $ & $2\times 14 $ & 0.85 & 0.55 & 0.90 & 0.75 & 0.82 & 0.12 \\
NGC 3628 & 9070080 & 2004-05-13 & $4\times 6 $\tablenotemark{d} & $4\times 6 $ & 1.72 & 1.38 & 1.66 & 0.95 & 1.00 & 0.38 \\
NGC 4088 & 9070336 & 2004-04-19 & $2\times 14 $ & $2\times 14 $ & 1.25 & 1.06 & 1.16 & 0.85 & 0.90 & 0.13 \\
NGC 4194 & 3757824 & 2004-01-08 & $2\times 14 $ & $4\times 6 $\tablenotemark{e} & 1.33 & 1.20 & 1.28 & 0.93 & 0.98 & 0.81 \\
NGC 4676 & 9073152 & 2004-05-15 & $6\times 14 $ & $4\times 30 $ & 1.30 & 1.08 & 1.23 & 0.95 & 0.97 & 0.69 \\
NGC 4818 & 9071104 & 2004-07-14 & $2\times 14 $ & $4\times 6 $\tablenotemark{e} & 1.05 & 0.95 & 1.03 & 0.98 & 0.98 & 0.81 \\
NGC 4945 & 8769280 & 2004-03-01 & $4\times 6 $  & $4\times 6 $  & 1.39 & 0.83 & 1.11 & 0.94 & 1.03 & 0.28 \\
NGC 7252 & 9074688 & 2004-05-15 & $3\times 14 $ & $2\times 30 $ & 1.54 & 1.18 & 1.38 & 0.95 & 0.98 & 0.90 \\
NGC 7714 & 3756800 & 2003-12-16 & $2\times 14 $ & $2\times 14 $ & 1.16 & 0.99 & 1.10 & 0.98 & 1.00 & 0.78 \\
\enddata
\label{tabobsprop}

\tablenotetext{a}{Exposure time in cycles times seconds. Each cycle in staring 
               mode corresponds to two exposures.  Hence, to derive the
               total exposure time one needs to multiply the above
               numbers by two.}
\tablenotetext{b}{Multiplicative factors for the SL1, SL2, SL3, LL2, 
               and LL3 modules, respectively, to stitch the spectral fragments 
               together with respect to LL1.  The unusually large stitching 
               factors used for NGC 520 are likely due to its very 
               irregular and extended structure, which led to a substantial 
               flux loss for the narrowest slits.}
\tablenotetext{c}{The fractional flux FF is the ratio of $25\mu$m flux detected
               within the LL slit to the total flux of the entire galaxy 
               measured by {\sl IRAS} \citep{san03}.} 
\tablenotetext{d}{SL2 was $2\times 14s$}
\tablenotetext{e}{LL2 was $2\times 14s$}
\end{deluxetable*}

\subsection{Data Reduction}

The data were pre-processed by the {\em Spitzer} Science Center (SSC)
data reduction pipeline version 11.0 \citep{som04} (except for
NGC\,3256, which was processed with version~12).  To avoid
uncertainties introduced by the flat-fielding in earlier versions of
the automated pipeline processing we started from the two-dimensional,
unflat-fielded data products, which only lack stray light correction
and flat fielding.  These products are part of the ``basic calibrated
data'' (BCD) package provided by the SSC.  The various steps of the
data reduction followed a recipe that has been developed by the {\sl
IRS} Disks team and tested on a large amount of Galactic and
extragalactic spectra.

We used the Spectral Modeling, Analysis, and Reduction Tool (SMART)
version 5.5.1., developed by the {\sl IRS} team \citep{hig04} to
reduce and extract the spectra.  First, we median-combined the images
of the same order, same module and same nod position. Then we
differenced the two apertures to subtract the ``sky'' background,
which is mainly from zodiacal light emission.  We have done so by
subtracting the spatially offset first and second order slits from
each other: SL1 from SL2 and vice versa for the SL (short wave, low
resolution) module, and LL1 from LL2 and vice versa for the LL (long
wave, low resolution) module.  We note that this approach assumes that
the emission from the target is not farther extended than the angular
distance between the two corresponding subslits of $79''$ in SL.

We extracted the spectra using a column width which increases
-- like the instrumental point-spread function -- linearly with
wavelength.  The extraction width is set to four pixels at the central
wavelength of each subslit.  
The spectra were flat-fielded and flux calibrated by
multiplication with the relative spectral response function (RSRF)
using the {\sl IRS} standard star $\alpha$-Lac for both low resolution
modules.  We built our RSRF by extracting the spectra of calibration
stars \citep{coh03} in the same way as we perform on our sources, then
divide the template spectra of those standard stars by the spectra we
extracted with the column extraction method. Finally, we multiplied
the spectra of the science target by the RSRF, for each nod position.

Figure~\ref{figslitoverlays} shows the positions of the narrower SL
and wider LL slits for the given observing date overplotted on mid-IR
images from IRAC, MIPS, and 2MASS in logarithmic scale.  Our
anticipated slit positions agree quite well with the main peak of the
mid-IR emission, except for NGC~1097, NGC~3310, and NGC~3556, which
show a more complex morphology. The consequences of slight mismatches
and extended emission are discussed in the following section.

\subsection{Absolute Fluxes and Order Stitching}
\label{secstitching}

After the spectral extraction there was, in some cases, a noticeable
mismatch between the spectra from the different {\sl IRS} modules.
Since this mismatch is more likely due to source flux that was missed
in the narrower slits rather than {\sl unrelated} flux that was picked
up in the wider slits (see below) we scaled the SL2, SL1, and LL2
spectra to match the flux density of LL1.  The choice of LL1 as
reference slit is appropriate because it has the widest slit and
largest PSF, and is least sensitive to pointing errors or a small
spatial extent of the mid-IR emission region.  We have also used the
``bonus orders'' SL3 and LL3 when they provided better overlap or
higher signal-to-noise than first and second orders only.  The applied
stitching factors are listed in Table~\ref{tabobsprop} and provide a
good idea of the uncertainties involved.  Since these factors are
rather large in some cases we would like to emphasize the rationale
for this approach.  The problem of stitching together slit apertures
of different widths is by no means specific to our approach, but
applies to basically all comparable studies at almost all wavelengths.
For a black-body-like object with extended, uniform surface brightness
there will be a jump between the SL slit width ($3.6''$) to the LL
slit $10.5''$, corresponding to an increase in flux of at least a
factor of three.  Since all spectral energy distributions are
continuous, scaling SL to match LL seems reasonable to first order.

However, this approach assumes that the spectral properties are not
changing within the region covered by the LL slit, corresponding to
linear scales of about 200\,pc for the nearest objects in our sample
-- the typical size of a circum-nuclear starburst over which the
spectral properties are assumed to not vary substantially.  If the
contributions from a centrally concentrated source, e.g., an AGN, were
dominant, scaling would lead to an overestimation of the strength of
the features originating in the center.  

For the latter reason we have decided to list the flux densities as
measured from the stitched spectra, but not to overall scale the {\sl
IRS} spectra to match the spatially integrated {\sl IRAS} flux
densities at $25\mu$m.  We have calculated the $\nu f_{\nu}$ average
of the {\sl IRS} spectra over the $25\mu$m {\sl IRAS} filter band.
The ratio of {\sl IRS} to {\sl IRAS} $25\mu$m flux densities is given
as the fractional flux 'FF' in Table~\ref{tabobsprop}, last column.  We
note that color corrections applied to the published {\sl IRAS}
catalog fluxes increase the uncertainties, but the relative effect on
our sample with similar SEDs is small.  In some cases the ratio is
quite small, indicating a rather large apparent discrepancy between
{\sl IRS} and {\sl IRAS}.  This is mainly due to two reasons: ({\sl
i}) the galaxy extends over a large angle and the mid-IR emission
region is significantly more extended than the {\sl IRS} slit (e.g.,
NGC~3628, NGC~4945), or ({\sl ii}) the galaxy has significant
off-nuclear IR emission peaks (e.g., NGC~3556, NGC~4088).

Figure~\ref{figslitoverlays} shows that substructure or extended
emission on scales of the {\sl IRS} slit is present in many of our
objects, most notably in those which have small fractional fluxes `FF'
in Table~\ref{tabobsprop}.  The most extreme cases are NGC~1097,
NGC~3556, and NGC~4088.  NGC~3556 shows a bright, off-nuclear mid-IR
source and several IR-bright knots along the disk.  NGC~4088 is very
extended with IR emission in the disk that is picked up in the large
IRAS beam.  In the case of NGC~1097 it is clear from
Figure~\ref{figslitoverlays} that the very symmetrical ring is the
reason why we only see 21\% of the flux.  Although we do not scale our
measured spectra to match the {\sl IRAS} fluxes, the factor FF will
become very important in section~\ref{secdiscus} where we use absolute
fluxes to derive total luminosities and star formation rates.

\begin{figure*}[tp]
\plotone{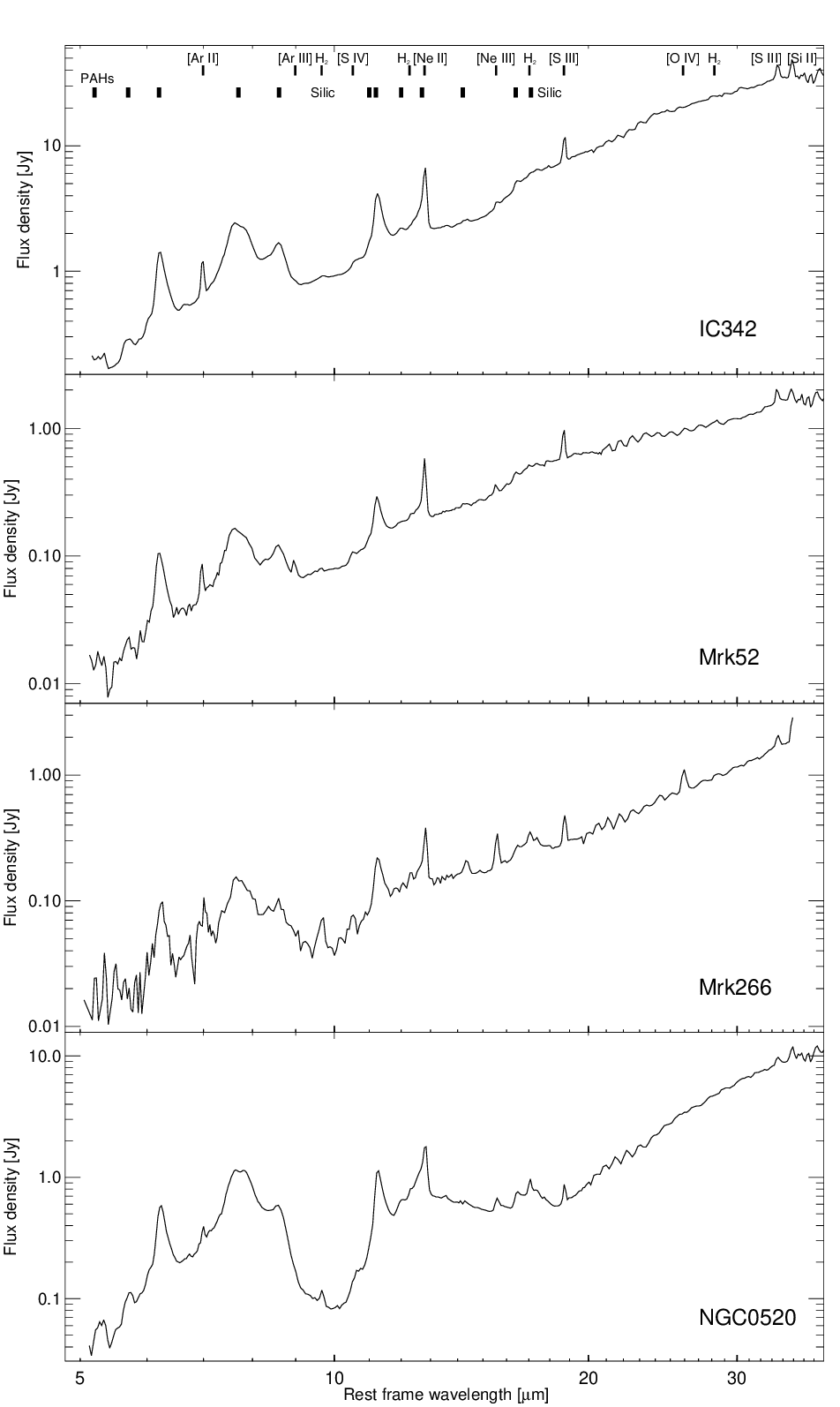}
\caption{{\sl Spitzer-IRS} low-resolution spectra of 
         of all starburts in our sample
         on logartithmic scale.  The most
         prominent spectral features (fine-structure lines, H$_2$
         lines, PAHs and silicate features) are labeled. For absolute
         flux calibration see the discussion in
         section~\ref{secstitching}.
         \label{fourspectra1}}
\end{figure*}

All of the individual {\sl IRS} spectra are shown in
Fig.~\ref{fourspectra1}.  We have not attempted to correct for the
periodic ``fringing'' in the spectra longward of about $22\mu$m, which
can be very prominent, as in NGC~1614.  These artifacts have no effect
on the analysis carried out in this paper.

\begin{figure*}[tp]
{\plotone{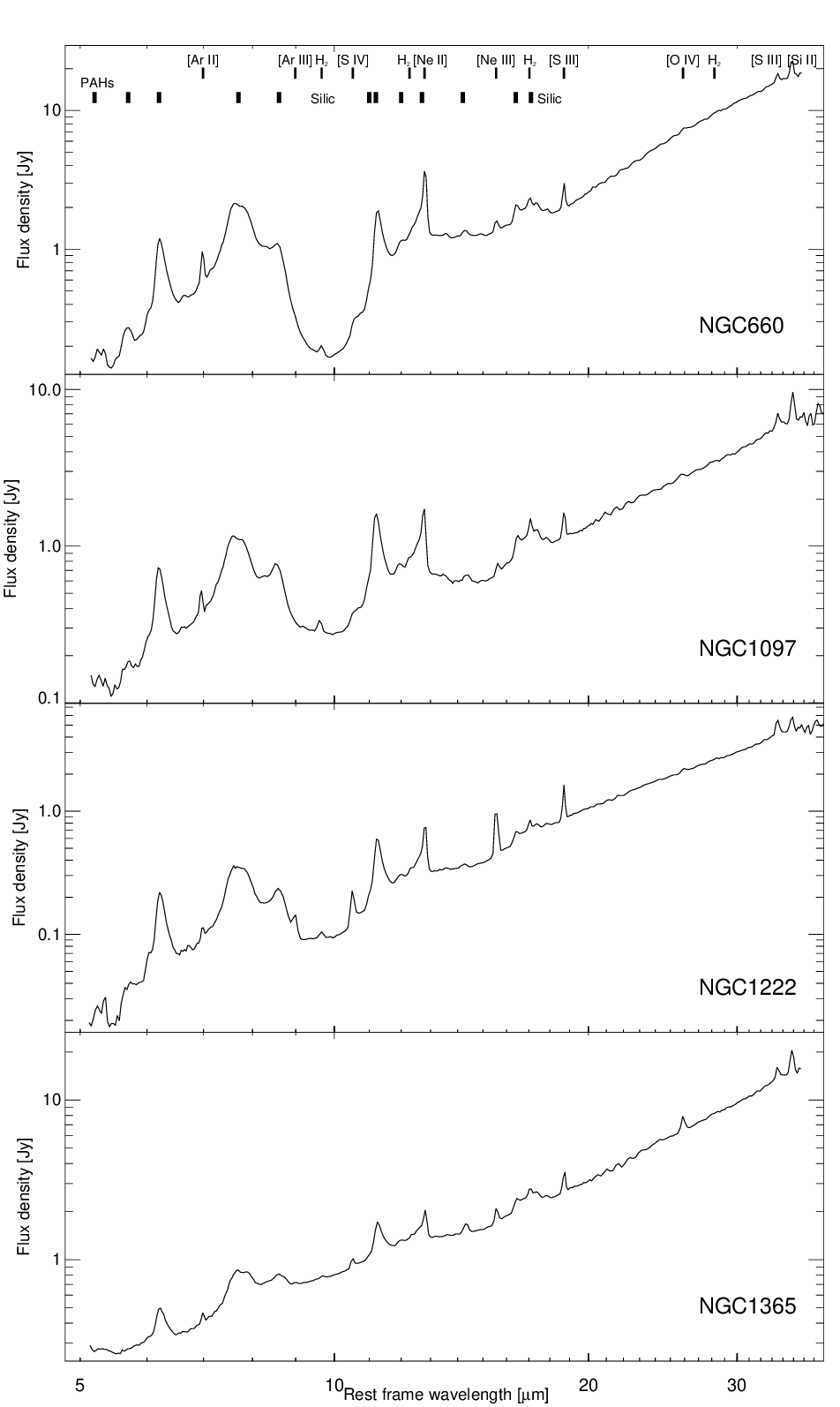}}\\
\centerline{{\sc Fig.~\ref{fourspectra1}.} cont'd --- }
\end{figure*}

\begin{figure*}[tp]
{\plotone{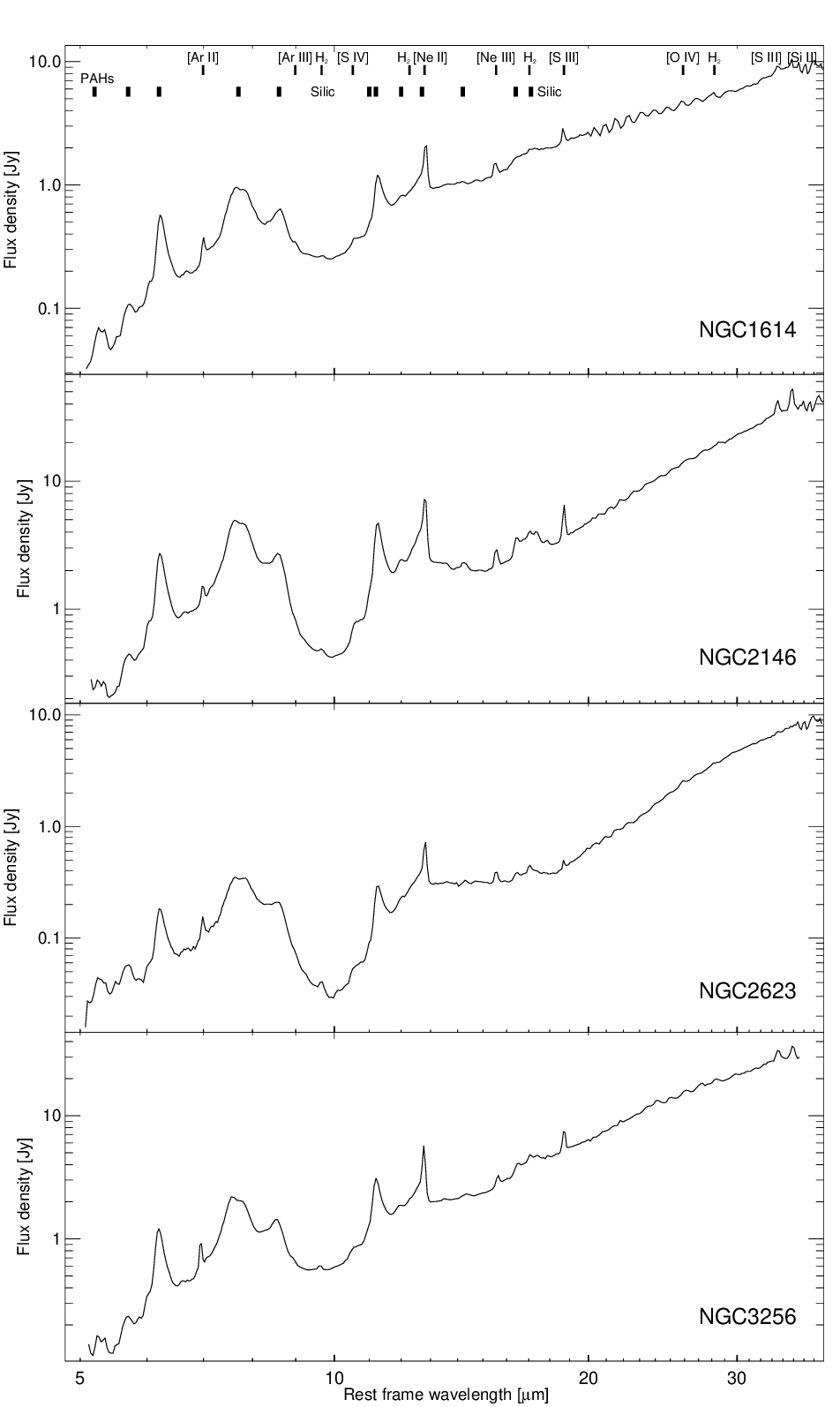}}\\
\centerline{{\sc Fig.~\ref{fourspectra1}.} cont'd ---}
\end{figure*}

\begin{figure*}[tp]
{\plotone{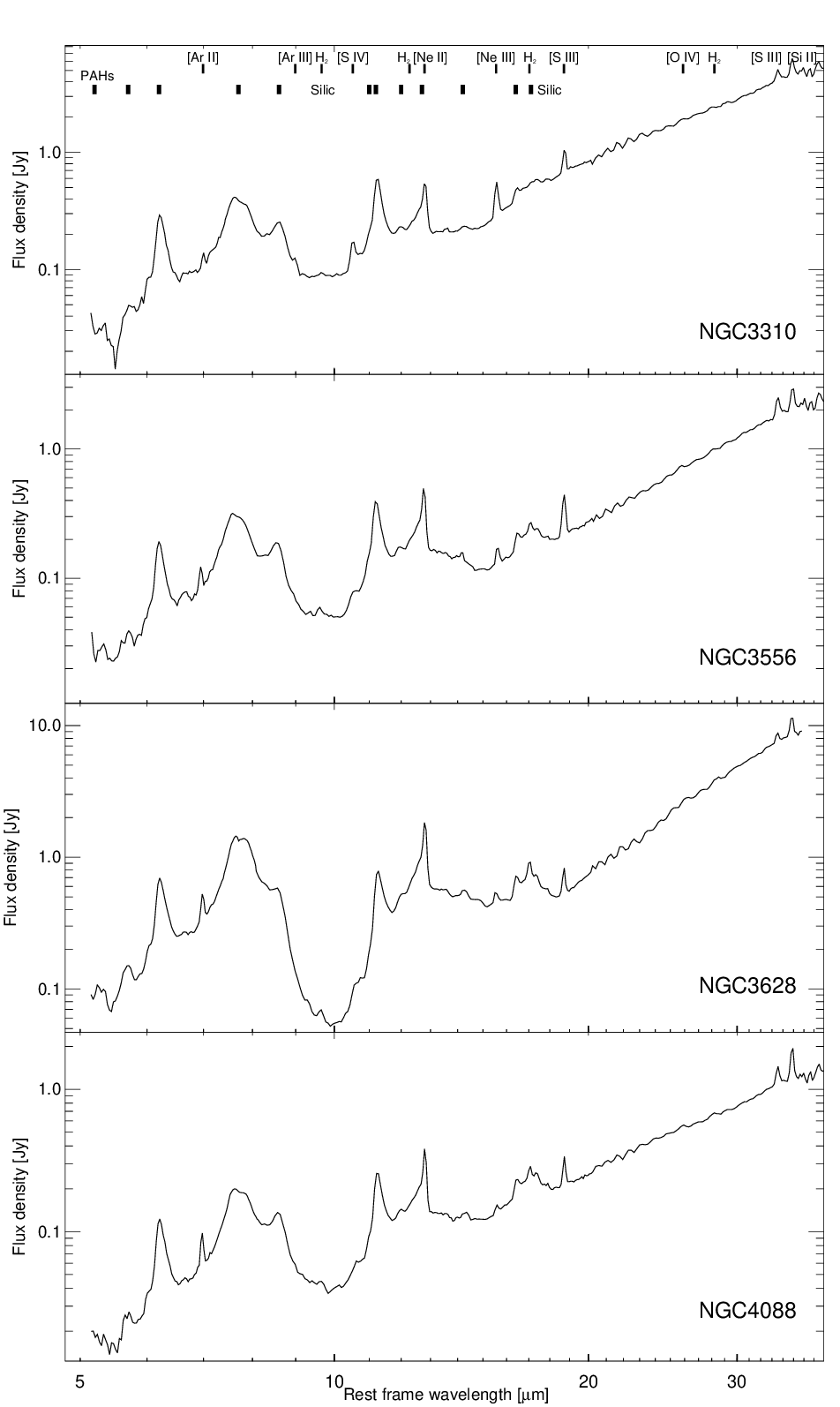}}\\
\centerline{{\sc Fig.~\ref{fourspectra1}.} cont'd --- }
\end{figure*}

\begin{figure*}[tp]
{\plotone{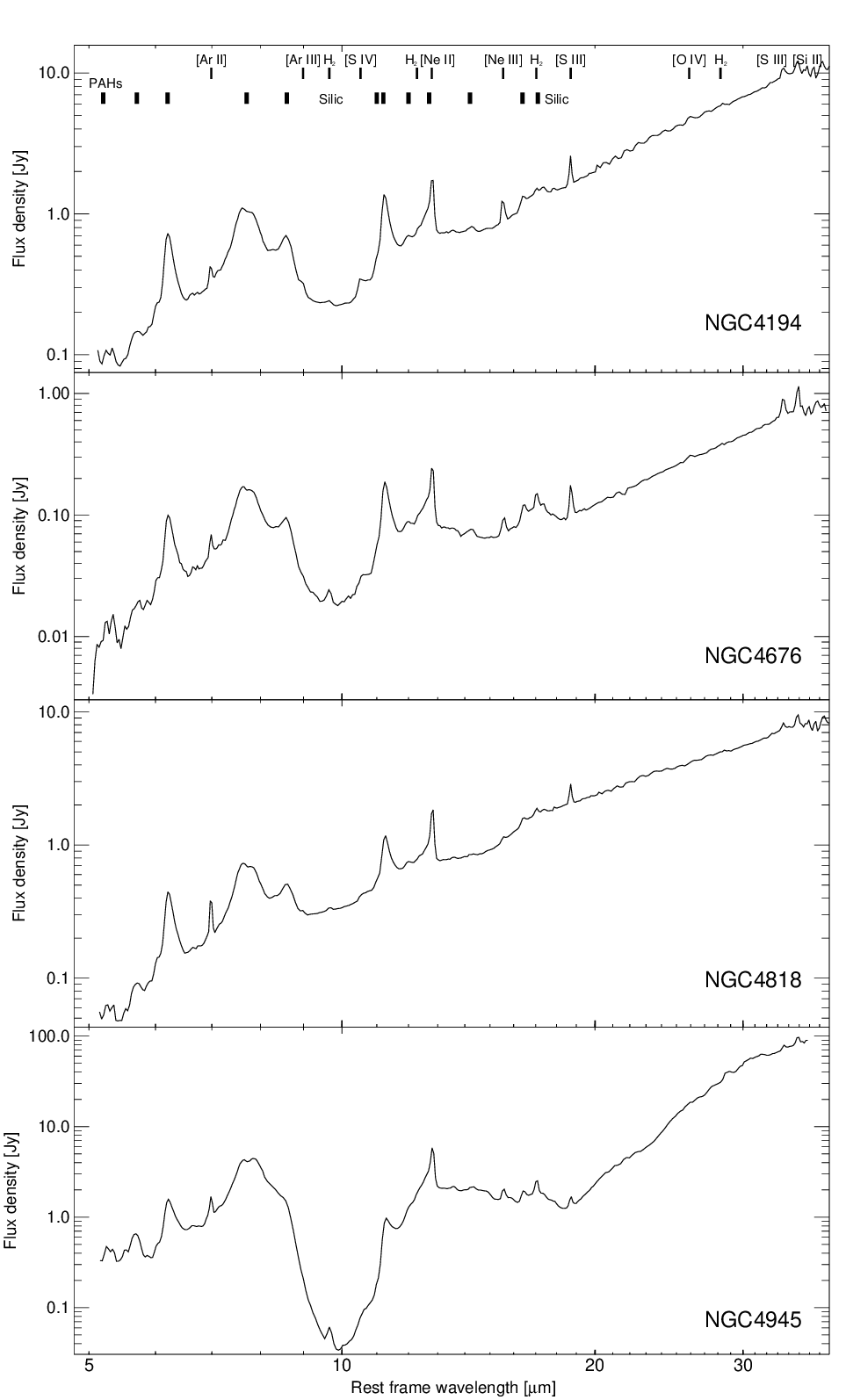}}\\
\centerline{{\sc Fig.~\ref{fourspectra1}.} cont'd --- }
\end{figure*}

\begin{figure*}[tp]
{\plotone{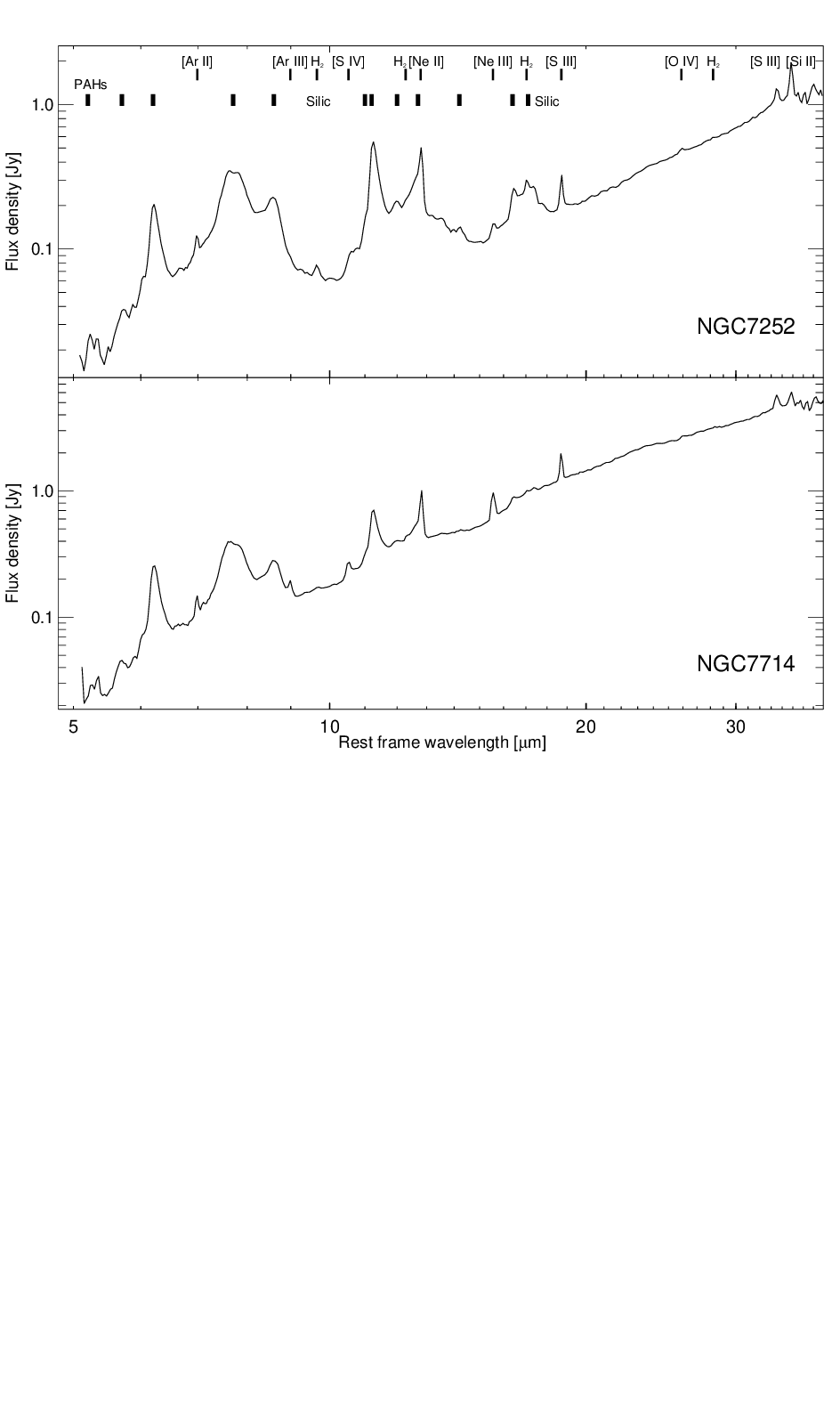}}\\
\centerline{{\sc Fig.~\ref{fourspectra1}.} cont'd --- }
\end{figure*}


\section{Analysis}
\label{secanaly}

\begin{figure*}[htb]
\epsscale{1.2}
\plotone{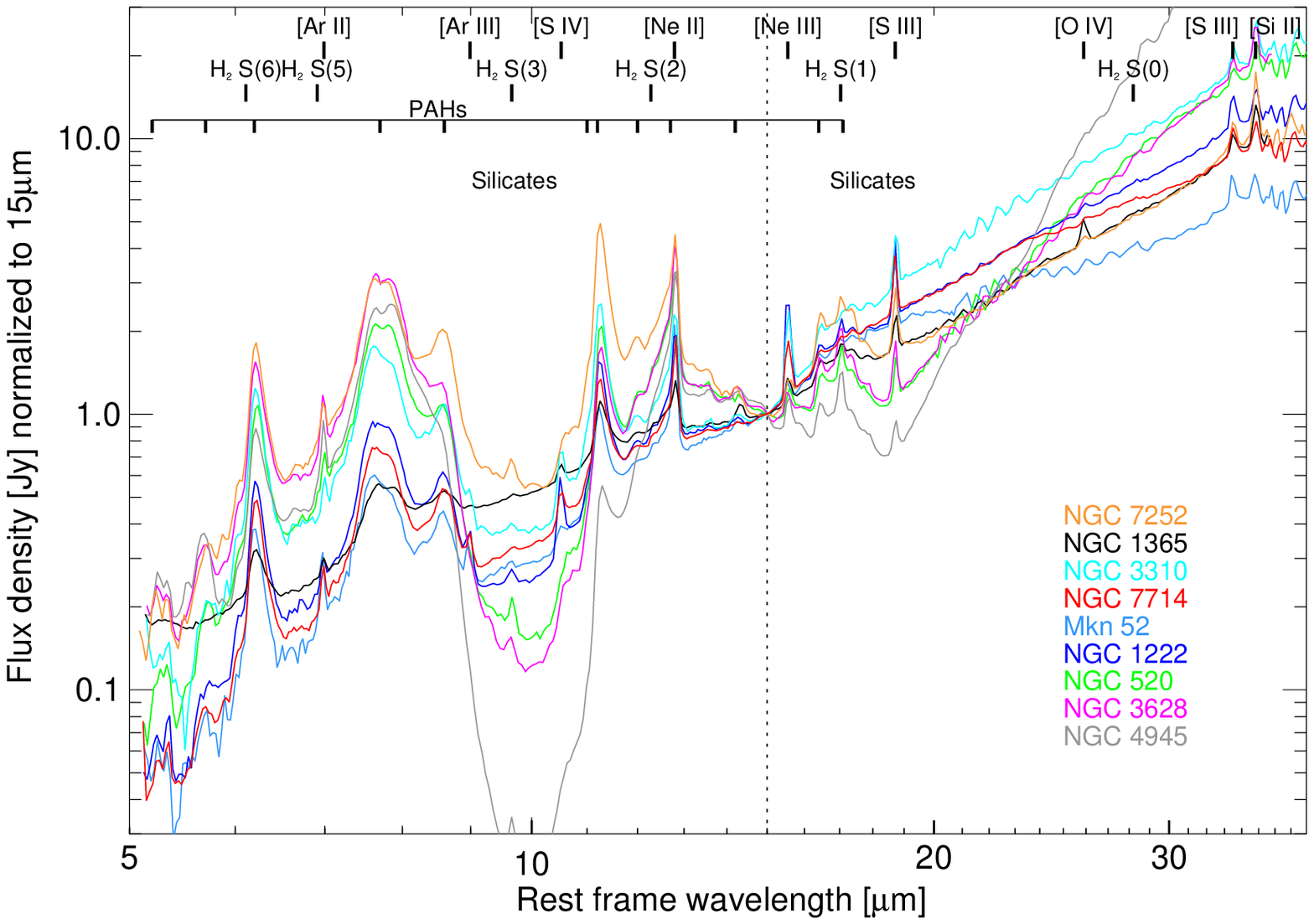}
\caption{Overlay of nine {\sl IRS} starburst spectra of Mkn 52, NGC
         520, NGC 1222, NGC 1365, NGC 3310, NGC 3628, NGC 4945, NGC
         7252, and NGC 7714. The objects shown here have been selected
         to illustrate the full spectral diversity of starbursts with
         a small number of objects.  All spectra have been normalized
         to a flux density of one at $15\mu$m (dotted vertical line).
         \label{figfinalall}}
\end{figure*}

Figure~\ref{figfinalall} shows a normalized overlay of nine starburst
spectra from our sample with the most prominant spectral features
labeled.  The figure illustrates the spectral richness of the $5 -
38\mu$m wavelength range and reveals distinct differences between
individual starbursts.  Important common features include PAH emission
bands, silicate emission or absorption features, and emission lines,
in addition to the information contained in the slope of the spectral
continuum.  In the following subsections we discuss how the quantities
relevant to our discussion have been measured from our spectra.

\subsection{Continuum Fluxes}
\label{seccont}

In order to characterize the basic properties of the spectral
continuum we have derived the flux densities for three rest frame
wavelength ranges: at $5.9 - 6.1\mu$m, at $14.75 - 15.25\mu$m, and at
$29.5 - 30.5\mu$m.  These wavelengths were chosen to cover a large
baseline in wavelength while being least affected by PAH emission
features, silicate absorption or strong emission lines.  The flux
densities derived for $6\mu$m, $15\mu$m, and $30\mu$m are the median
values in the above wavelength ranges, respectively, and are arguably
the best direct estimate of the spectral continuum.
The measured fluxes are listed in Table~\ref{tabspecfits}.  In
section~\ref{secbrandllumi} we will use these continuum fluxes to
estimate the total luminosity of the starburst galaxy.

\begin{deluxetable}{l r r r r}
\centering 
\tabletypesize{\footnotesize} 
\tablecaption{Continuum fluxes and extinction} 
\tablewidth{0pt} 
\tablehead{
\colhead{Name} & 
\colhead{$F_{6\mu m}$} & \colhead{$F_{15\mu m}$} &
\colhead{$F_{30\mu m}$} & \colhead{$\tau_{9.8}$}\\
  & \colhead{[Jy]} & \colhead{[Jy]} & \colhead{[Jy]} & }
\startdata 
IC   342 & 0.38 & 2.71 & 27.26 &  0.004 \\
Mrk   52 & 0.03 & 0.28 &  1.19 & -0.003 \\
Mrk  266 & 0.03 & 0.17 &  1.16 &  0.373 \\
NGC  520 & 0.15 & 0.54 &  6.12 &  0.994 \\
NGC  660 & 0.34 & 1.27 & 11.52 &  1.293 \\
NGC 1097 & 0.12 & 0.29 &  3.23 &  0.130 \\
NGC 1222 & 0.06 & 0.38 &  3.03 &  0.213 \\
NGC 1365 & 0.32 & 1.55 &  9.61 & -0.034 \\
NGC 1614 & 0.14 & 1.10 &  5.80 &  0.279 \\
NGC 2146 & 0.74 & 2.00 & 23.07 &  0.845 \\
NGC 2623 & 0.06 & 0.32 &  4.75 &  1.544 \\
NGC 3256 & 0.34 & 2.36 & 21.61 &  0.000 \\ 
NGC 3310 & 0.08 & 0.24 &  2.79 &  0.057 \\
NGC 3556 & 0.02 & 0.10 &  1.10 &  0.230 \\
NGC 3628 & 0.19 & 0.46 &  4.89 &  1.640 \\
NGC 4088 & 0.04 & 0.12 &  0.76 &  0.307 \\
NGC 4194 & 0.22 & 0.79 &  6.77 &  0.371 \\
NGC 4676 & 0.03 & 0.07 &  0.45 &  0.580 \\
NGC 4818 & 0.13 & 0.92 &  5.56 & -0.160 \\
NGC 4945 & 0.48 & 1.83 & 49.56 &  4.684 \\
NGC 7252 & 0.06 & 0.11 &  0.69 & -0.098 \\
NGC 7714 & 0.07 & 0.53 &  3.48 & -0.096 \\
\enddata
\label{tabspecfits}
\end{deluxetable}

\subsection{Polycyclic Aromatic Hydrocarbons (PAHs)}
\label{pahanalys}

The spectra in Fig.~\ref{figfinalall} show that the spectral continuum
shape of starburst spectra is dominated by strong emission features
from PAHs (as previously noted by {\sl ISO} authors, e.g.,
\citet{gen00} and references therein).  Although the first detections
of PAHs date back in the 1970s, it took more than 10 years to identify
them.  Here we concentrate our analysis on several of the strongest
PAH features at $6.2\mu$m, $7.7\mu$m, $8.6\mu$m, $11.3\mu$m,
$14.2\mu$m, and around $17\mu$m (which is in fact two blended PAH
complexes centered at $16.4\mu$m and $17.1\mu$m).  Further PAH
features can be seen in the spectra (see sections~\ref{secices} and
\ref{secpahvari}) but have either low S/N or are, at the low
resolution of the {\sl IRS} SL+LL modules, blended with other features
and were thus excluded from this analysis.  At the {\sl IRS}
low-resolution the $12.7\mu$m PAH blends with the strong [Ne~II] line
at $12.81\mu$m.  This feature will be discussed by \citet{dev06} in
the context of the {\sl IRS} high resolution spectra.

\begin{figure}[htb]
\includegraphics[angle=90,width=9cm]{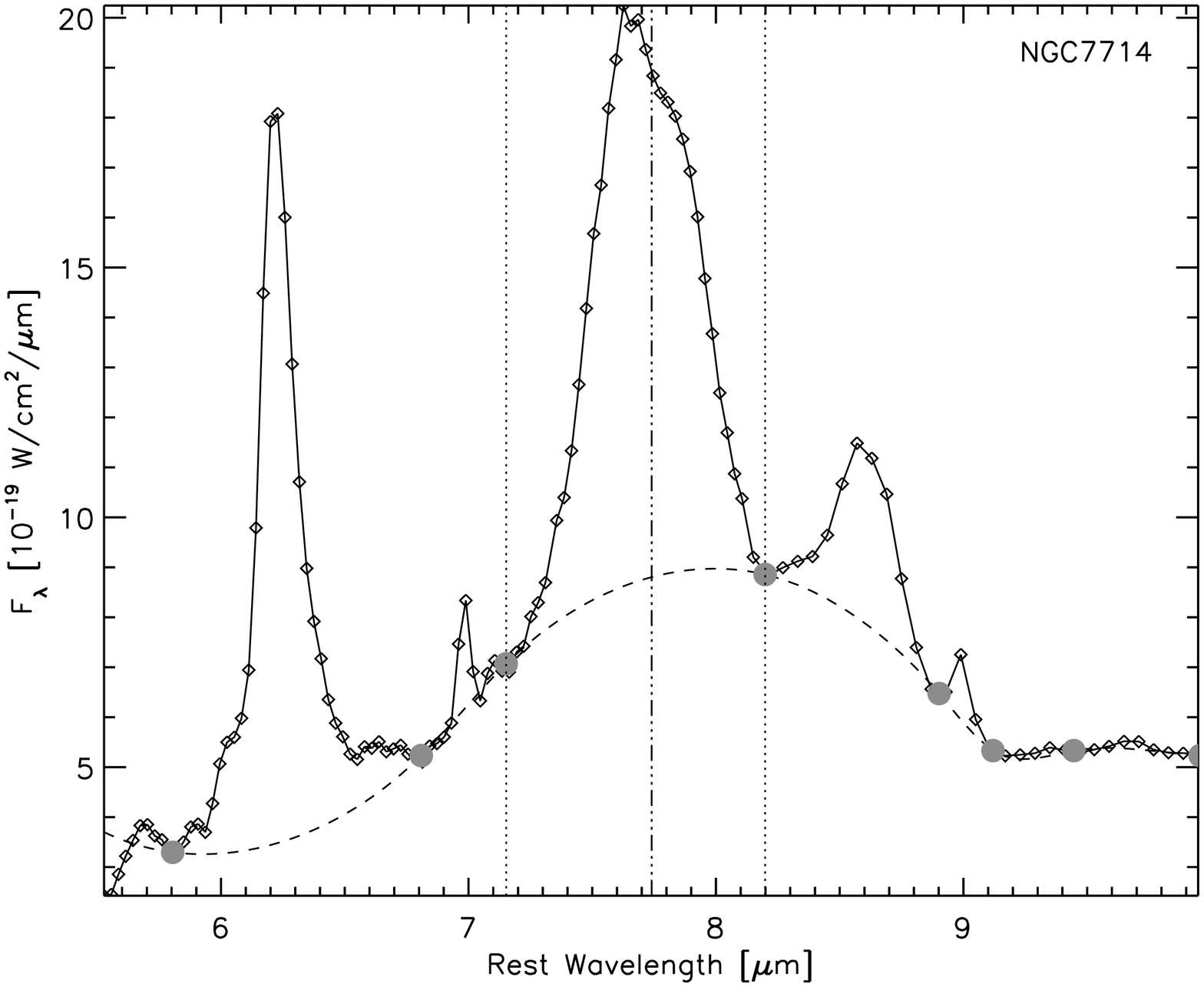}
\caption{Illustration of how the $7.7\mu$m PAH strength was measured
         in NGC~7714.  The light grey dots represent the continuum
         points, interpolated by the dashed spline function. The
         vertical dotted lines indicate the integration range for the
         $7.7\mu$m feature.\label{figpahfit}}
\end{figure}

The strengths of the $6.2\mu$m, $7.7\mu$m, $8.6\mu$m, and $11.3\mu$m
PAH emission bands were derived by integrating the flux of the feature
in the mean spectra of both nod positions above an adopted continuum.
For the 6.2 and $11.3\mu$m features this baseline was determined by
fitting a spline function to four or five data points. The wavelength
limits for the integration of the features were approximately between
5.94 to $6.56\mu$m in the case of the $6.2\mu$m PAH and between 10.82
and $11.80\mu$m for the $11.2\mu$m PAH.  For most spectra this method
produced results, reproducable to within 5\% for repeated fits with
different choices of the continuum or integration limits to account
for uncertainties within the fitting procedure.  Only for the noisiest
spectra this uncertainty increased to 15\%.  The baseline for the the
$7.7\mu$m and $8.6\mu$m features was derived by fitting a spline
through six data points avoiding small features in the range between
5.5 and $10\mu$m.  Our method is illustrated in
Figure~\ref{figpahfit}. The selected data points for the baseline were
chosen at the same wavelengths for all spectra.  Following
\citet{pee02} we included one point close to $8.2\mu$m to separate the
contribution from both features.

The PAH features at $14.2\mu$m \citep{hon01} and $17\mu$m
\citep{ker00} are intrinsically weaker but are located in spectrally
less complex regions, allowing a different approach.  The strength of
the $14.2\mu$m feature was determined through a first order baseline
fit to the continuum at $13.88-14.03\mu$m and $14.54-14.74\mu$m.
Similarly, the strength of the $17\mu$m complex was determined through
a first order baseline fit to the continuum at $15.95-16.15\mu$m and
$18.20-18.40\mu$m.  We emphasize that the feature we refer to as the
$17\mu$m PAH is in fact a blend of two PAH complexes at $16.4\mu$m and
$17.1\mu$m.  The latter is furthemore contaminated by the H$_2$S(1)
line at $17.0\mu$m.  The individual components of this complex cannot
be properly resolved in the low-resolution spectra and we give only
combined fluxes here.  The equivalent widths for all features were
derived by dividing the integrated PAH flux above the adopted
continuum by the continuum flux density at the center of the feature
(indicated by the vertical dot-dashed line in Figure~\ref{figpahfit}).
The derived PAH fluxes and equivalent widths are listed in
Table~\ref{tabpahs}.

\begin{deluxetable*}{l r r r r r r r r r r r r}
\centering 
\tabletypesize{\footnotesize} 
\tablecaption{Main PAH feature strengths shortward of $18\mu$m}
\tablewidth{0pt} 
\tablehead{ 
\colhead{} & 
\multicolumn{2}{c}{$6.2\mu$m PAH} & 
\multicolumn{2}{c}{$7.7\mu$m PAH} & 
\multicolumn{2}{c}{$8.6\mu$m PAH} & 
\multicolumn{2}{c}{$11.3\mu$m PAH} &
\multicolumn{2}{c}{$14.2\mu$m PAH} &
\multicolumn{2}{c}{$17\mu$m PAH} \\
\colhead{Name} &
\colhead{Flux\tablenotemark{a}} & \colhead{EW\tablenotemark{b}} &
\colhead{Flux\tablenotemark{a}} & \colhead{EW\tablenotemark{b}} &
\colhead{Flux\tablenotemark{a}} & \colhead{EW\tablenotemark{b}} &
\colhead{Flux\tablenotemark{a}} & \colhead{EW\tablenotemark{b}} &
\colhead{Flux\tablenotemark{a}} & \colhead{EW\tablenotemark{b}} &
\colhead{Flux\tablenotemark{a}} & \colhead{EW\tablenotemark{b}}
}
\startdata 
 IC342 & 14.65 & 0.497 & 31.75 & 0.581 & 7.39 & 0.168 & 18.36 & 0.492 & 0.50 & 0.014 & 9.70 & 0.178  \\  
 Mrk52 & 1.07 & 0.535 & 2.14 & 0.552 & 0.51 & 0.151 & 1.02 & 0.316 & 0.02 & 0.005 & 0.66 & 0.142  \\  
 Mrk266 & 0.59 & 0.619 & 0.92 & 0.467 & 0.00 & 0.126 & 0.47 & 0.422 & 0.08 & 0.065 & 0.51 & 0.403  \\  
 NGC0520 & 5.60 & 0.563 & 13.97 & 0.528 & 1.99 & 0.126 & 5.62 & 0.798 & 0.20 & 0.024 & 2.91 & 0.494  \\  
 NGC0660 & 12.91 & 0.504 & 27.36 & 0.518 & 3.81 & 0.123 & 9.72 & 0.701 & 0.43 & 0.023 & 6.26 & 0.365  \\  
 NGC1097 & 5.04 & 0.459 & 9.30 & 0.488 & 2.26 & 0.168 & 5.34 & 0.661 & 0.12 & 0.021 & 3.82 & 0.523  \\  
 NGC1222 & 2.17 & 0.624 & 4.64 & 0.606 & 0.79 & 0.130 & 2.56 & 0.566 & 0.08 & 0.015 & 1.47 & 0.231  \\  
 NGC1365 & 2.74 & 0.111 & 6.37 & 0.213 & 1.30 & 0.045 & 4.60 & 0.180 & 0.62 & 0.029 & 5.59 & 0.250  \\  
 NGC1614 & 13.05 & 0.561 & 24.67 & 0.514 & 5.01 & 0.138 & 9.77 & 0.379 & --- & --- & 5.10 & 0.154  \\  
 NGC2146 & 22.22 & 0.545 & 58.26 & 0.643 & 10.51 & 0.175 & 21.53 & 0.829 & 1.20 & 0.044 & 13.60 & 0.492  \\  
 NGC2623 & 1.83 & 0.598 & 4.00 & 0.454 & 0.76 & 0.131 & 1.28 & 0.527 & 0.01 & 0.002 & 0.71 & 0.205  \\  
 NGC3256 & 8.02 & 0.603 & 17.60 & 0.533 & 3.03 & 0.123 & 8.52 & 0.471 & 0.71 & 0.040 & 8.12 & 0.205  \\  
 NGC3310 & 3.35 & 0.789 & 5.38 & 0.591 & 1.16 & 0.178 & 2.98 & 0.748 & 0.07 & 0.022 & 1.08 & 0.229  \\  
 NGC3556 & 1.22 & 0.502 & 2.92 & 0.523 & 0.51 & 0.135 & 1.52 & 0.811 & 0.06 & 0.039 & 0.84 & 0.542  \\  
 NGC3628 & 7.45 & 0.500 & 19.72 & 0.588 & 1.59 & 0.095 & 4.14 & 0.797 & 0.25 & 0.034 & 3.45 & 0.684  \\  
 NGC4088 & 1.17 & 0.496 & 2.43 & 0.483 & 0.51 & 0.130 & 1.27 & 0.603 & 0.05 & 0.025 & 0.97 & 0.509  \\  
 NGC4194 & 7.09 & 0.529 & 14.67 & 0.578 & 3.06 & 0.165 & 6.31 & 0.590 & 0.25 & 0.022 & 3.10 & 0.241  \\  
 NGC4676 & 1.05 & 0.610 & 2.15 & 0.551 & 0.44 & 0.192 & 0.97 & 0.812 & 0.04 & 0.038 & 0.52 & 0.590  \\  
 NGC4818 & 4.06 & 0.459 & 9.56 & 0.555 & 1.84 & 0.123 & 4.39 & 0.344 & 0.05 & 0.004 & 2.43 & 0.151  \\  
 NGC4945 & 12.13 & 0.432 & 40.34 & 0.490 & 0.13 & 0.003 & 3.55 & 0.558 & 0.50 & 0.024 & 5.37 & 0.519  \\  
 NGC7252 & 2.09 & 0.585 & 4.48 & 0.549 & 1.06 & 0.176 & 2.95 & 0.931 & 0.04 & 0.024 & 1.35 & 0.792  \\  
 NGC7714 & 2.67 & 0.601 & 5.62 & 0.642 & 1.06 & 0.135 & 2.79 & 0.394 & 0.04 & 0.006 & 1.05 & 0.114  \\  
\enddata
\label{tabpahs}
\tablenotetext{a}{Flux in units of $10^{-19}$W\,cm$^{-2}$}
\tablenotetext{b}{Equivalent width in units of $\mu$m}
\end{deluxetable*}

It is important to note that the values in Table~\ref{tabpahs} have
not been corrected for extinction.  While in many sources the silicate
absorption band around $9.8\mu$m is very weak, objects severely
affected by extinction, like NGC~4945, show a strong $7.7\mu$m PAH but
much weaker $8.6\mu$m and $11.3\mu$m features.  The cause of these
variations will be addressed in section~\ref{secpahvari}.

We do not list uncertainties for the PAH strengths in
Table~\ref{tabpahs}.  The statistical errors from the fits to the high
S/N are small compared to other uncertainties such as:
{\sl (i)} The error in the absolute flux calibration of each module,
which is currently about $5-10\%$.
{\sl (ii)} The error from scaling the individual orders for an
extended source to match, as discussed in
section~\ref{secstitching}.
{\sl (iii)} The error in the approximation of the underlying
continuum, which is often dominated by strong, adjacent emission and
absorption features and varies from source to source, and 
{\sl (iv)} The error in the strength of the $17\mu$m PAH from the
blending of two PAH complexes and the H$_2$S(1) line.

Our measurement procedures have been designed to minimize these errors
as best as possible.  The dominant error remains the uncertainty in the
flux calibration, and we estimate the error in the PAH measurements to
be in the order of $10\%$.

\subsection{Emission Lines}
\label{seclines}

The $5 - 38\mu$m wavelength range contains numerous strong emission
lines.  Among those are the following forbidden lines -- sorted by
wavelength and with their excitation potentials in parentheses:

\noindent
[Ar\,II]\,$6.99\mu$m, 
[Ar\,III]\,$8.99\mu$m,
[S\,IV]\,$10.51\mu$m,
[Ne\,II]\,$12.81\mu$m,
[Ne\,III]\,$15.56\mu$m,
[S\,III]\,$18.71\mu$m,
[O\,IV]\,$25.89\mu$m,
[S\,III]\,$33.48\mu$m, and
[Si\,II]\,$34.82\mu$m.
In addition, we detect the pure rotational lines of molecular hydrogen
H$_2$\,(0,0)\,S(5)\,$6.91\mu$m (blended with [Ar\,II]), 
H$_2$\,(0,0)\,S(3)\,$9.66\mu$m, 
H$_2$\,(0,0)\,S(2)\,$12.28\mu$m, and
H$_2$\,(0,0)\,S(1)\,$17.03\mu$m.

All of these lines have been identified and labeled in
Fig.~\ref{figfinalall}.  We list them here since they can be easily
detected, even at $R \leq 100$.  However, the flux measurements of the
fine-structure lines can be done much more accurately from the {\sl
IRS} high-resolution spectra, which is the subject of a complementary
paper discussing the ionic properties of the ISM \citep{dev06}.

\subsection{Silicate Absorption and Optical Depth}
\label{secsili}

The wavelength coverage of {\sl IRS} is ideally suited for a detailed
study of the strong vibrational resonances in the silicate mineral
component of interstellar dust grains.  Amorphous silicates --- the
most common form of silicates --- have a broad Si--O stretching
resonance peaking at $9.8\mu$m and an even broader O--Si--O bending
mode resonance peaking at $18.5\mu$m.

We have estimated the apparent optical depth in the $9.8\mu$m silicate
feature from the ratio of the local mid-infrared continuum to the
observed flux at $9.8\mu$m. For shallow silicate features, the local
continuum may be defined as an F$_{\lambda}$ power law interpolation
between continuum pivots at $5.5\mu$m (averaged $5.3-5.7\mu$m flux)
and $14.5\mu$m (averaged $14.0-15.0\mu$m flux), thus avoiding the main
PAH emission complexes at $6-9\mu$m and $11-13\mu$m.  

For starburst spectra with a more pronounced silicate feature, the
continuum in the $14-15\mu$m range is affected by weak absorption from
the overlapping wings of the 9.8 and $18\mu$m silicate features.  For
these spectra we define a second local continuum, by replacing the
continuum pivot at $14.5\mu$m by a continuum pivot at $28\mu$m
(averaged $27.5-28.5\mu$m flux) and use the average of the optical
depths derived with either local continuum as the best estimate of the
apparent $9.8\mu$m silicate optical depth.


Most fluxes used in this analysis are observed flux densities.
However, in Fig.~\ref{figpahrationeon} we correct the observed fluxes
for extinction.  To get an estimate of the uncertainties we use two
extinction laws from \citet{dra89} and \citet{lut99} and show the
difference in Fig.~\ref{figpahrationeon}.  The relative silicate
absorption values $A_{\lambda} / A_V$ for the relevant PAH wavelengths
are listed in Table~\ref{tabextlaws}.

\notetoeditor{Place table~\ref{tabextlaws} here.}
\begin{deluxetable}{r c c}
\centering 
\tabletypesize{\footnotesize} 
\tablecaption{Relative absorption values interpolated from \citet{dra89} 
              and \citet{lut99}} 
\tablewidth{0pt} 
\tablehead{
\colhead{Wavelength} & 
\colhead{$A_{\lambda}^{Draine} / A_V$} & 
\colhead{$A_{\lambda}^{Lutz} / A_V$}
}
\startdata 
$6.2\mu$m  & 0.0164 & 0.0489 \\
$7.7\mu$m  & 0.0112 & 0.0440 \\
$9.8\mu$m  & 0.0554 & 0.1289 \\
$11.3\mu$m & 0.0375 & 0.0872 \\
\enddata
\label{tabextlaws}
\end{deluxetable}

Some estimates of $\tau_{9.8\mu m}$ in Table~\ref{tabspecfits} are
negative, implying that silicates are observed in emission.
However, the absolute values are quite small and may just represent
uncertainties in our baseline definition (see section~\ref{variety}
for a discussion).  We also note that the true silicate optical depth
may be significantly larger than the apparent silicate optical depth
if the emitting and absorbing sources are mixed along the line of
sight; if part of the silicate column is warm; or if the absorption
spectrum is diluted by unrelated foreground emission.


\subsection{Spectral Features in the $5-8\mu$m Range}
\label{secices}

The $5-8\mu$m spectral range of starburst galaxies is extremely rich
in atomic and molecular emission and absorption features, and
dominated by emission from the $6.2\mu$m PAH feature and the blue wing
of the $7.7\mu$m PAH complex. Weaker emission features are expected
from atomic lines ([Fe\,II] at $5.34\mu$m and [Ar\,II] at $6.99\mu$m),
molecular hydrogen (H$_2$ S(7) at $5.51\mu$m and H$_2$ S(5) at
$6.91\mu$m), and `combination-mode' PAH emission bands (at $5.25\mu$m
and $5.70\mu$m).  Absorption features of water ice and hydrocarbons,
commonly detected in deeply obscured galactic nuclei \citep{spo02},
would be expected at $6.0\mu$m (water ice) and $6.90\mu$m and
$7.25\mu$m (C--H bending modes in aliphatic hydrocarbons).

\begin{figure}[ht]
\epsscale{1.2}
\plotone{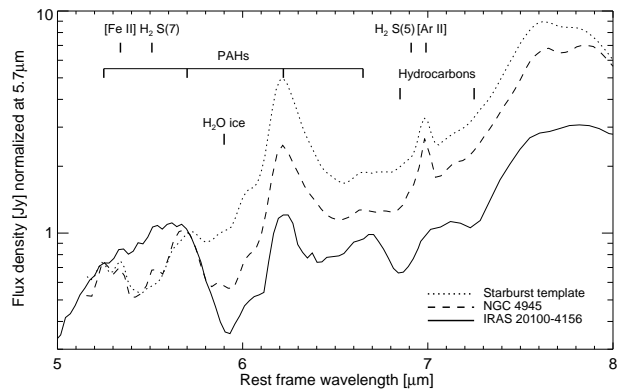}
\caption{Comparison of the average starburst template (dotted line)
         from Fig.~\ref{figaverage} to the most extincted source
         within our sample, NGC~4945 (dashed line), and to the ULIRG
         IRAS~20100-4156 (solid line).  The spectra are normalized at
         $5.7\mu$m.  IRAS~20100-4156 has been chosen as a
         representative ULIRG from the sample of \citet{spo06}.
         \label{figiceabsorb}}
\end{figure}

\begin{figure}[htb]
\epsscale{1.2}
\plotone{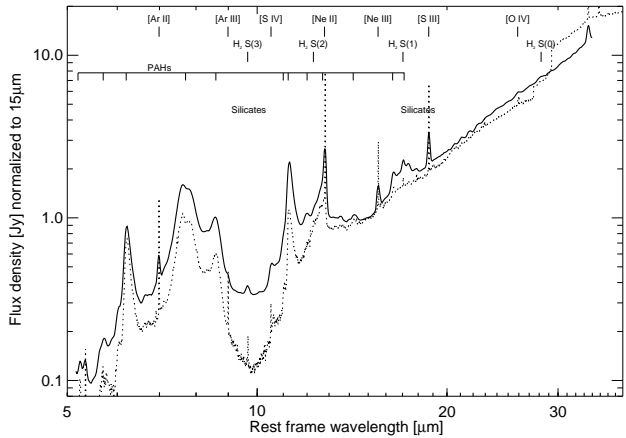}
\caption{Average {\sl IRS} spectrum of 13 starburst galaxies (IC 342,
         NGC 660, NGC 1097, NGC 1222, NGC 2146, NGC 3310, NGC 3556,
         NGC 4088, NGC 4194, NGC 4676, NGC 4818, NGC 7252, and NGC
         7714).  All spectra have been normalized to a flux density of
         one at $15\mu$m before co-addition.  The dotted line shows
         the ISO-SWS spectrum of M82 \citep{stu00} for comparison. 
	 \label{figaverage}}
\end{figure}

As illustrated by the average starburst spectrum in
Fig.~\ref{figaverage}, PAH combination-mode emission features are
common in our starburst spectra. Inspection of the individual spectra
shows that the $5.25\mu$m feature is usually double-peaked due to
blending with the $5.34\mu$m [Fe\,II] line (e.g. NGC~1222 and
NGC~3256).  Likewise, the profile of the $5.70\mu$m PAH feature is
affected by the presence of the H$_2$ S(7) line at $5.51\mu$m (most
notably NGC~2623 and NGC~4945). The $5.70\mu$m PAH feature appears
strongest in the spectrum of NGC~4945 (Fig.~\ref{figiceabsorb}).  The
ratio of the $5.7\mu$m to $6.2\mu$m PAH in this source is 0.29, about
5 times higher than for most other starburst galaxies in our sample.
Interestingly, the red wing of the $5.7\mu$m PAH feature coincides
with the steep onset of the $6\mu$m water ice absorption feature, as
illustrated in Fig.~\ref{figiceabsorb} by the steep change in slope at
$5.7\mu$m in the spectrum of the ULIRG IRAS~20100--4156. Simple
spectral modeling confirms that a screen of water ice absorption can
indeed mimick a stronger $5.7\mu$m PAH feature by suppressing the red
wing of the feature and the adjacent $5.9\mu$m continuum. The presence
of water ice in the nucleus of NGC\,4945 is further supported by the
discovery of a $3\mu$m water ice absorption feature in the ISO--PHT--S
and VLT--ISAAC spectra of the nucleus \citep{spo00}.  For the
remaining galaxies in our sample, the $5-6\mu$m spectral structure
does not provide evidence for the presence of water ice
absorption. Hence, apart from NGC~4945, shielded cold molecular gas
may not be as abundant in starburst nuclei as in ULIRG nuclei.

Absorption features of aliphatic hydrocarbons at 6.85 and $7.25\mu$m
are thought to be tracers of the diffuse ISM \citep{chi00}. These
features are readily detected in the spectra of deeply obscured ULIRG
nuclei such as IRAS\,20100--4156 (Fig.~\ref{figiceabsorb}).  The average
starburst spectrum (Fig.~\ref{figaverage}), in contrast, does not show
similarly pronounced structure and the $7\mu$m range is dominated
instead by the blend of $6.91\mu$m H$_2$ S(5) and $6.99\mu$m [Ar\,II].
However, individual starburst spectra show weak spectral structure in
the $6.5-7.0\mu$m range.  At $6.65\mu$m, a weak emission feature seems
to be present, most notably in the spectra of IC~342, NGC~660,
NGC~1614, NGC~2146, NGC~4088, NGC~4945 and NGC~7252. \citet{pee02}
identify this feature as a PAH emission band.  The red wing of the
$6.65\mu$m emission feature lies close to the expected onset of the
$6.85\mu$m hydrocarbon absorption feature
(Fig.~\ref{figiceabsorb}). Especially in the spectrum of NGC~4945, the
$6.5-7.0\mu$m spectral structure is consistent with the presence of
hydrocarbon absorption at a strength of
$\tau$(6.85\,$\mu$m)=0.15$\pm$0.05 (Fig.~\ref{figiceabsorb}). For other
starburst galaxies the spectral structure is too shallow and/or the
S/N of the spectra too low to identify hydrocarbon absorption with
sufficient confidence.

\subsection{The Starburst ``Template Spectrum''}

Although many spectral features show variations from one starburst
galaxy to another, the magnitude of these changes is relatively small
compared to the differences between different classes of objects, such
as AGN, quasars, ULIRGs, or normal galaxies.  It is common practice to
classify objects in these categories, and the availability of
reference spectra is of great interest.  A reference ``template''
spectrum would allow a comparison how close a given spectrum is to a
typical starburst galaxy or if it shows any atypical features.
Furthermore, classifications of objects at high redshift often require
fits to a library of template spectra.

We have constructed a high signal-to-noise ``starburst template'' from
the spectra of IC 342, NGC 660, NGC 1097, NGC 1222, NGC 2146, NGC
3310, NGC 3556, NGC 4088, NGC 4194, NGC 4676, NGC 4818, NGC 7252, and
NGC 7714.  These are basically the objects from our sample with high
fluxes and without a strong AGN component.  The spectra have been
normalized to a flux density of unity at $15\mu$m before averaging.
Figure~\ref{figaverage} shows the resulting ``template spectrum'' and
Table~\ref{tabsbaverage} lists the ``average'' spectral properties
derived from this composed spectrum.

\begin{deluxetable}{l l}
\centering 
\tabletypesize{\footnotesize} 
\tablecaption{Properties of the ``average'' starburst galaxy} 
\tablewidth{0pt} 
\tablehead{
\colhead{Parameter} & \colhead{Value}
}
\startdata 
$F_{6\mu m}$  & 26\% \\
$F_{15\mu m}$ & 100\% \\
$F_{30\mu m}$ & 856\% \\ 
$6.2\mu$m  PAH EW & $0.53\mu$m \\ 
$7.7\mu$m  PAH EW & $0.53\mu$m \\
$8.6\mu$m  PAH EW & $0.15\mu$m \\
$11.3\mu$m PAH EW & $0.66\mu$m \\
$14.2\mu$m PAH EW & $0.02\mu$m \\
$17\mu$m   PAH EW & $0.36\mu$m \\
$\tau_{9.8}$ & $0.24\pm 0.10$\\
\enddata
\label{tabsbaverage}
\end{deluxetable}

Figure~\ref{figaverage} also shows the ISO-SWS spectrum of M82
\citep{stu00}, which is often being used as a starburst template, for
comparison.  Although the spectral slope longward of $15\mu$m is very
similar, the two spectra show several distinct differences: the
ISO-SWS spectrum of M82 does not show the pronounced PAH complex
around $17\mu$m, it shows much stronger silicate absorption,
and the flux density shortward of $12\mu$m is almost a factor of two
lower that in our average starburst template.  We provide the spectrum
in ascii table format on our website\footnote{\tt
http://www.strw.leidenuniv.nl/~brandl/SB\b{ }template.html}.


\section{Results and Discussion} 
\label{secdiscus}

It is important to keep in mind that the starburst spectra presented
in this paper represent an entire starburst region, including numerous
(super-)star clusters at various ages and evolutionary states, the
surrounding PDRs which are internally and externally excited, and the
warm and cold dust spread across the entire region as well as
localized dust condensations.  While many of the properties of local
substructures are expected to vary significantly, the overall
significance of these variations may be averaged out in the spatially
integrated spectra.
Our aim here is to search for global trends between the spectral
properties derived in the previous section (silicate absorption
features, PAH features, spectral continuum) and the global starburst
properties ($L_{IR}$, radiation field).


\subsection{The Continuum Slope as a SB/AGN Diagnostic}
\label{seccontflux}

\begin{figure}[ht]
\epsscale{1.2}
\plotone{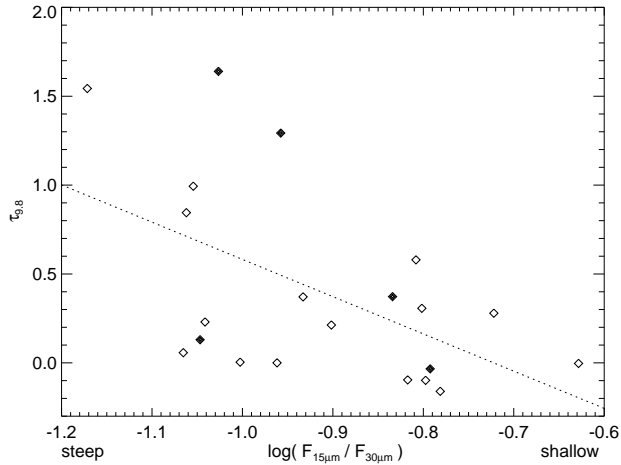}
\caption{Optical depth at $9.8\mu$m versus the slope of the continuum
         as measured by the flux ratio of $15\mu$m to $30\mu$m.  The
         filled diamonds correspond to starbursts with a weak AGN
         component.  The dotted line is a linear fit to the data
         points. The highly obscured source NGC~4945 is not included
         in the fit..
	 \label{figtauslope}}
\end{figure}

The slope of the mid-IR spectral continuum depends on the optical
thickness, composition and temperature of graphite dust grains, which
are related to the amount of silicate grains \citep{mat77}.  The
dominating species in the {\sl IRS} spectral range are hot ($\ge
100$\,K), large grains heated by ionizing, non-ionizing, and
Ly$\alpha$ photons inside \2 regions, and small ($\le 100$\,\AA)
grains heated by non-ionizing photons outside \2 region \citep{mou97}.
In Fig.~\ref{figtauslope} we plot the continuum slope, parametrized by
the ratio of the $15\mu$m and $30\mu$m flux densities versus the
optical depth at $9.8\mu$m (section~\ref{secsili}).  The filled
symbols in Fig.~\ref{figtauslope} (and all other figures thereafter)
correspond to starbursts with a weak AGN component as identified in
Table~\ref{tabgenprop}.  Although not a tight correlation one can see
the general trend that starbursts with stronger silicate absorption
tend to have a steeper continuum at longer wavelengths.

The slope of the dust continuum depends on the energy distribution and
spatial concentration of the heating source(s).  \citet{dal00} found
from {\sl ISOCAM} data of 61 galaxies a $6.75\mu$m\,/\,$15\mu$m
continuum slope near unity for more quiescent galaxies, whereas that
ratio drops (i.e., the slope steepens) for galaxies with increased
starburst activity.  Furthermore, the continuum slope can serve as a
discriminator between massive stars or an AGN as the underlying power
source.  This has already been known since {\sl IRAS} (e.g.,
\citet{wan92}), and was further refined in numerous papers based on
{\sl ISOCAM} observations. (See \citet{gen00} for a more comprehensive
overview). For instance, \citet{lau00} studied the {\sl ISOCAM} colors
of a large variety of extragalactic objects revealing clear general
trends between AGN, photo-dissociation region (PDR), and \2 region
dominated spectra.  However, the characterization of an individual
object often turns out to be difficult due to the intrinsically large
scatter: the {\sl IRAS} $12\mu$m filter includes the silicate
absorption band as well as several PAH emission features and strong
emission lines.  The wavelength range covered by {\sl ISOCAM} is
limited to shorter wavelengths which are dominated by hot dust and a
large variety of emission and absorption features (compare to
Fig.~\ref{figiceabsorb}).

\begin{figure}[ht]
\epsscale{1.2}
\plotone{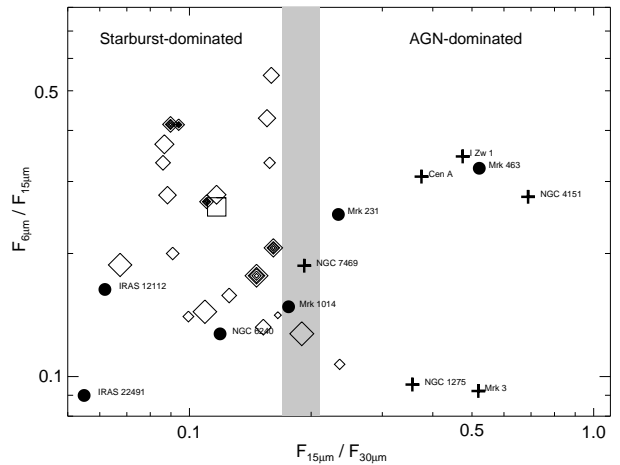}
\caption{``Color-color plot'' of the $6\mu$m/$15\mu$m versus
         $15\mu$m/$30\mu$m flux densities.  The size of the diamonds,
         representing our starburst sample, increases linearly with
         $\log L_{IR}$.  The filled diamonds correspond to starbursts
         with a weak AGN component and the square refers to the
         average starburst template (Table~\ref{tabsbaverage}).  For
         comparison the figure also contains AGN (pluses) from
         \citet{wee05}, as well as ULIRGs (filled circles) from
         \citet{arm06a} and \citet{arm06b}.
\label{figcontfluxes}}
\end{figure}

Our narrowband continuum fluxes largely avoid these problems.
Figure~\ref{figcontfluxes}, analogous to a color-color diagram in
stellar astronomy, shows the continuum flux ratios at $6\mu$m/$15\mu$m
versus $15\mu$m/$30\mu$m.  The total infrared luminosity $L_{IR}$ is
represented by the size of the symbols.  For comparison the figure
also contains the classical AGN Cen\,A (Sy\,2), I\,Zw\,1 (Sy\,1),
Mrk~3 (Sy\,2), NGC~1275 (Sy\,2), NGC~4151 (Sy\,1.5), and NGC~7469
(Sy\,1.2) from \citet{wee05}.  We also show the starburst-dominated
ULIRGs IRAS~12112+0305, IRAS~22491-1808, and the AGN-dominated
Mrk~231, Mrk~463, and Mrk~1014 from \citet{arm06b}.  NGC~6240 is a
peculiar case with an intrinsic fractional AGN contribution to the
bolometric luminosity of $~20-24\%$ \citep{arm06a}.

Several conclusions can be drawn from Fig.~\ref{figcontfluxes}.
First, there is a large scatter along the $y$-axis with no obvious
correlation with total starburst luminosity or galaxy type.  Hence,
the $6\mu$m/$15\mu$m continuum flux ratio does not appear to be a good
diagnostic.  Second, using the $15\mu$m/$30\mu$m continuum flux ratio,
classical (strong) AGN can be clearly separated from the starburst
galaxies (including the ones with weak nuclear activity), with the AGN
having a significantly shallower mid-IR spectrum.  The grey-shaded bar
in Fig.~\ref{figcontfluxes} indicates the transition region between
AGN and starburst-dominated systems, and lies approximately at
\begin{displaymath}
0.17\le F_{15\mu m}/F_{30\mu m} \le 0.21.
\end{displaymath}
Third, the difference in spectral slope also seems to apply to ULIRGs
depending on their dominant power source.  Hence, this technique may
have a much broader application and should be verified with a much
larger sample of different classes of objects.


\subsection{The Continuum Fluxes as Measures of $L_{IR}$}
\label{secbrandllumi}

The total infrared luminosity is an important parameter to estimate
the energetics of a starburst and to characterize the underlying
stellar population and rate of star formation. $L_{IR}$ is often
derived from the four {\sl IRAS} filter bands \citep{san96}. In this
subsection we will check how accurately $L_{IR}$ can be derived from
the two {\sl IRS} continuum fluxes at $F_{15\mu m}$ and $F_{30\mu m}$
alone.

Numerous attempts to extrapolate $L_{IR}$ from one or two, mainly
broad band fluxes can be found in the literature. \citet{tak05}
discuss various estimators of infrared luminosities, and find -- for a
very large sample of 1420 galaxies of different type -- correlations
of the form $\log L_{IR} = 1.02 + 0.972 \log L_{12\mu m}$, and $\log
L_{IR} = 2.01 + 0.878 \log L_{25\mu m}$.  Both $L_{12\mu m}$ and
$L_{25\mu m}$ allow to predict $L_{IR}$ to an accuracy within a factor
of $4-5$ at the 95\% confidence level over a wide range in
luminosities.  These uncertainties are similar to the luminosities
derived from the {\sl MIPS} $24\mu$m flux alone for a large sample of
{\sl SINGS} galaxies \citep{dal05}.  \citet{elb01} found a similar
relation fitting $15\mu$m {\sl ISO} fluxes: $\log L_{IR} = (1.05\pm
0.174) + 0.998\log L_{15\mu m}$.  \citet{for04} found that the
monochromatic $15\mu$m continuum emission is directly proportional to
the ionizing photon luminosity, and hence the total infrared
luminosity.

\begin{figure}[ht]
\epsscale{1.2}
\plotone{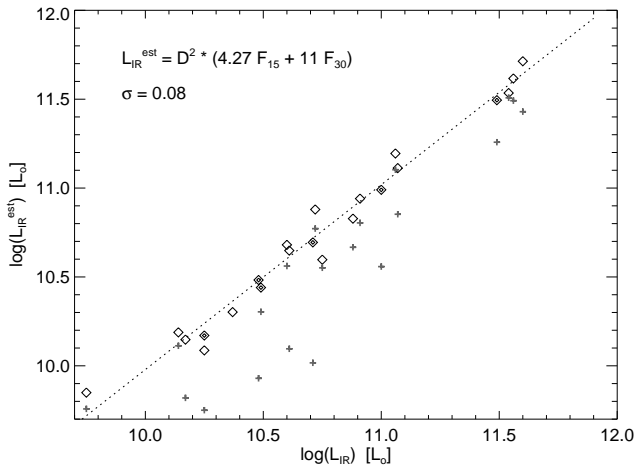}
\caption{The estimated infrared luminosity $L_{IR}^{est}$ based on the
         aperture loss corrected {\sl IRS} continuum fluxes $F_{15\mu
         m}$ and $F_{30\mu m}$ for a source at distance $D$ [kpc]
         versus $L_{IR}$ from {\sl IRAS}.  The dashed line represents
         a linear fit to the data (in log-log space), and the fitted
         function is given in the upper left, together with the
         standard error $\sigma$.  The uncorrected fluxes are
         indicated by gray plus signs.  The filled symbols correspond to
         starbursts with a weak AGN component. 
\label{figbrandlluminosity}}
\end{figure}

As discussed in section~\ref{seccontflux} the {\sl IRS} fluxes
$F_{15\mu m}$ and $F_{30\mu m}$ provide a rather accurate estimate of
the ``true'' spectral continuum.  In Figure~\ref{figbrandlluminosity}
we plot a combination of $F_{15\mu m}$ and $F_{30\mu m}$ times $D^2$
versus the total infrared luminosity derived from the {\sl IRAS}
bands.  Since the {\sl IRAS} beam usually covers the entire starburst
while the narrower {\sl IRS} slits can only collect a fraction of the
total luminosity for local starbursts (Fig.~\ref{figslitoverlays}) we
have corrected the observed $F_{15\mu m}$ and $F_{30\mu m}$ (indicated
by the pluses in Fig.~\ref{figbrandlluminosity}) for the slit losses
(see section~\ref{secstitching} for more details).  A least-square fit
to the corrected fluxes yields:
\begin{displaymath}
  L_{IR}^{est} = D^2\cdot \left(4.27\cdot F_{15\mu m} + 11\cdot F_{30\mu m}
  \right)
\end{displaymath}
where $D$ is the distance in kiloparsecs, $F_{15\mu m}$ and $F_{30\mu
m}$ the {\sl IRS} flux densities in Janskies.

The correlation in Fig.~\ref{figbrandlluminosity} is extremely tight,
including the weak AGN, with a standard error (mean scatter) of only
0.09 in log-space, i.e., the {\sl IRS} estimated infrared luminosities
agree within 23\% with $L_{IR}$.  This is much more accurate
than the estimates by \citet{elb01} and \citet{tak05}.  The excellent
correlation suggests that, at least for a homogeneous sample of
starburst galaxies, $F_{15\mu m}$ and $F_{30\mu m}$ can be
used to accurately derive $L_{IR}$.


\subsection{A Large Variety in Silicate Absorption}
\label{variety}

\begin{figure}[htb]
\epsscale{1.2}
\plotone{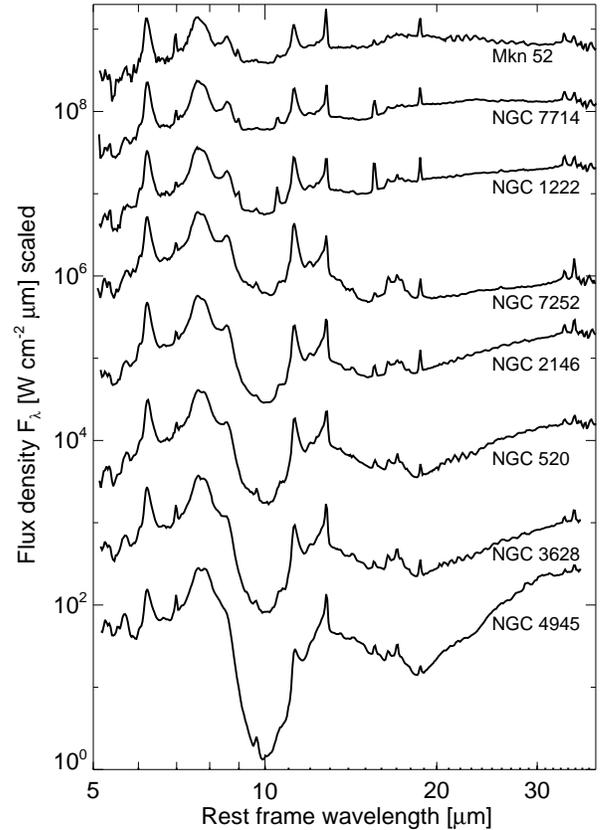}
\caption{The {\sl IRS} spectra of Mkn~52, NGC~7714, NGC~1222, NGC~7252,
         NGC~2146, NGC~520, NGC~3628, and NGC~4945 arranged to
         illustrate the gradual effect of increasing silicate
         absorption from top to bottom.  The flux densities
         $F_{\lambda}$ have been arbitrarily scaled for better
         comparision. \label{figoneontop}}
\end{figure}

Figure~\ref{figoneontop} shows, from top to bottom, a series of
starburst spectra, sorted by increasing absorption of the $9.8\mu$m
silicate resonance.  Since the peak of the resonance coincides with a
minimum between the $7-9\mu$m and $11-13\mu$m PAH emission complexes,
the effect of silicate absorption only becomes apparent towards the
lower half of the plot.  Nevertheless, the figure strikingly
illustrates the strong effect of amorphous silicates on the overall
$5-38\mu$m spectral shape and the large variations even within one
class of objects.  Fig.~\ref{figoneontop} shows that the $10\mu$m
trough can in fact become a dominating feature of the spectral shape
of starbursts.

The top two spectra also show a broad emission structure beginning at
about $16\mu$m and extending to $21-25\mu$m. The shape of this feature
is consistent with that of an $18\mu$m silicate emission feature,
which would make these the first detections of this feature in
starburst galaxies.  However, the emission structure could also be due
to the C--C--C in-plane and out-of-plane bending modes of PAHs
\citep{ker00}.  The identification of this feature will be discussed
in a future paper.  The remaining starburst galaxies in
Fig.~\ref{figoneontop} do not show evidence for a similar emission
structure. The $13-35\mu$m spectra, from NGC~1222 (top) to NGC~4945 at
the bottom, show an increasingly pronounced depression, peaking at
$18.5\mu$m, signalling increasingly strong silicate absorption. The
latter result is in full agreement with the trend found for the
$9.8\mu$m silicate feature.

Among all the galaxies classified as starbursts, NGC~4945 (bottom
spectrum in Fig.~\ref{figoneontop}) is a ``special case'' exhibiting
by far the strongest dust obscuration to its nuclear region
\citep{spo00}.  Based on the IRS spectrum, the apparent optical depth
in the $9.8\mu$m silicate feature is at least four and may be higher,
depending on the choice of the local continuum.  Apart from strong
amorphous silicate absorption, the line of sight also reveals the
presence of a $23\mu$m absorption feature.  Following the analysis of
deeply obscured lines of sight towards ULIRG nuclei \citep{spo06}, we
attribute the $23\mu$m feature to crystalline silicates (forsterite).
Their detection in NGC~4945 suggests that crystalline silicates are
perhaps a more common component of the ISM and not just limited to
ULIRG nuclei.

\begin{figure}[ht]
\epsscale{1.2}
\plotone{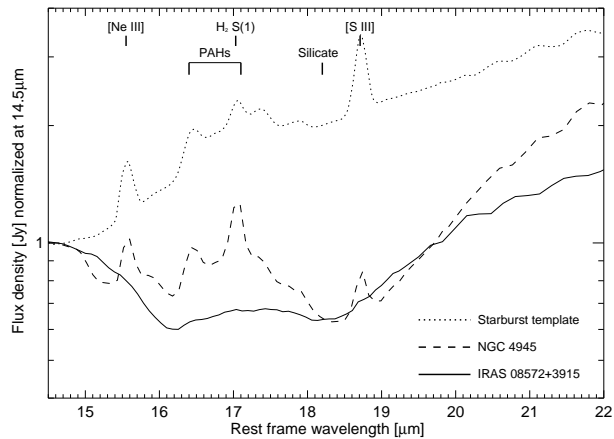}
\caption{Comparison of the average starburst template (dotted line)
         from Fig.~\ref{figaverage} to the most extincted source
         within our sample, NGC~4945 (dashed line), and to the ULIRG
         IRAS~08572+3915 (solid line), chosen as an extreme case of
         extinction from the sample of \citet{spo06}.  The spectra
         have been normalized at $14.5\mu$m.
         \label{figsilicabsorb}}
\end{figure}

Figure~\ref{figsilicabsorb} compares the average starburst template
from Fig.~\ref{figaverage} to NGC~4945, the most extincted source
within our sample.  For comparison we also show the heavily embedded
ULIRG IRAS~08572+3915 from \citet{spo06}.  The usually rather shallow
$18\mu$m silicate band reduces the continuum by a large factor
compared to the average starburst spectrum.  While NGC~4945 has a
similarly strong silicate absorption feature as the ULIRG
IRAS~08572+3915, the starburst also shows strong PAH complexes near
$16.4\mu$m and $17.1\mu$m, which are absent in the ULIRG spectrum.
Crystalline silicates, causing the features at 16 and $19\mu$m in
IRAS~08572+3915, appear to be absent in the starburst template,
although they are more difficult to discern at our spectral resolution
given the presence of PAH features and the [S\,III] line in that
range.

While NGC~4945 shows by far the most extreme silicate absorption
within our sample it is not an exotic object but defines the endpoint
of a sequence of increasing optical depths.  Recently, \citet{dal06}
reported on the {\sl Spitzer-SINGS} survey of 75 nearby galaxies.  The
{\sl typical} target in their sample shows only modest dust
obscuration, consistent with an $A_V\sim 1$ foreground screen and a
lack of dense clumps of highly obscured gas and dust.  Unfortunately,
we cannot distinguish between the local and global dust distributions.
The differences may be due to evolutionary states, local geometries or
other reasons.  However, it is evident that starburst galaxies can
have very little or large amounts of extinction, and the
presence/absence of a strong dust feature is not a characteristic
item.


\subsection{The Mixture of PAHs and Dust}
\label{secpahmixture}

In this section we will investigate how the derived PAH strengths may
depend on extinction within the starburst region.  Table~\ref{tabpahs}
already indicates that the relative fluxes of individual PAH features
are not constant for different starbursts.  \citet{lu03} have found a
25\% spread in the ratio of the $11.3\mu$m\,/\,$7.7\mu$m PAH fluxes
which they attribute to intrinsic galaxy-to-galaxy variations.
However, extinction can affect the relative strength of features at
different wavelengths by different amounts (Table~\ref{tabextlaws}).
\citet{rig99} found that the dominant influence on the (ULIRG) PAH
ratio is extinction, and that the $6.2\mu$m PAH gets suppressed
relative to the $7.7\mu$m PAH for heavily dust extincted systems.  In
particular, the $8.6\mu$m and $11.3\mu$m PAH features lie at
wavelengths that are heavily affected by silicate absorption.

\begin{figure}[ht]
\epsscale{1.2}
\plotone{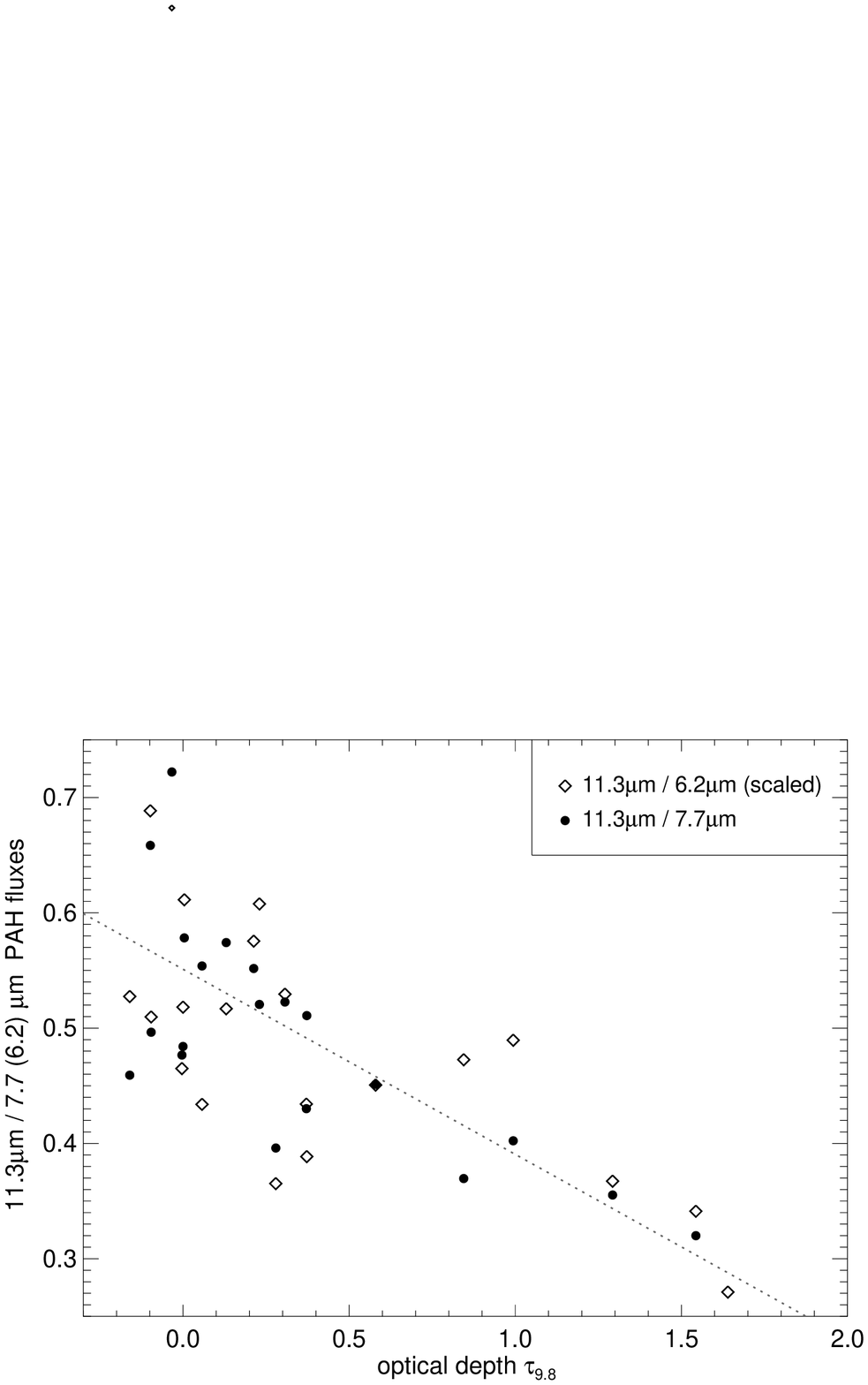}
\caption{The PAH flux ratios $11.3\mu$m/$7.7\mu$m (diamonds) and
         $11.3\mu$m/$6.2\mu$m (filled circles, divided by a factor
         2.05 to match the other ratio) versus the optical depth at
         $9.8\mu$m, $\tau_{9.8}$ from Table~\ref{tabspecfits}.  The
         dashed line shows a linear fit to both ratios but excluding
         NGC~4945. \label{fig11pahextinction}}
\end{figure}

Fig.~\ref{fig11pahextinction} shows the $11.3\mu$m\,/\,$7.7\mu$m and
$11.3\mu$m\,/\,$6.2\mu$m PAH flux ratios versus the apparent silicate
optical depth $\tau_{9.8}$.  A trend that starbursts with stronger
dust absorption show relatively weaker $11.3\mu$m PAHs is evident.
Fig.~\ref{fig11pahextinction} suggests that extinction can affect the
relative PAH strength in starbursts by up to a factor of about two.

\begin{figure}[ht]
\epsscale{1.2}
\plotone{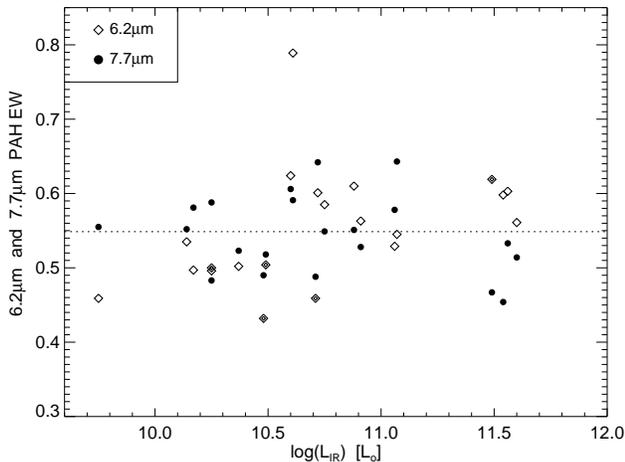}
\caption{The $6.2\mu$m (diamonds) and $7.7\mu$m (filled circles) PAH
         equivalent widths versus infrared luminosity $L_{IR}$.  The
         dashed, horizontal line indicates a zeroth order polynomial
         fit to all data points, excluding the outlier NGC~1365.  The
         filled diamonds correspond to starbursts with an AGN
         component.\label{figpahews}}
\end{figure}

In the remainder of this subsection we will investigate if the PAH
equivalent width is related to the total infrared luminosity of the
starburst.  \citet{rig99} found a ratio of $L_{PAH} / L_{IR}$ which is
similar for starburst-dominated ULIRGs and for template starbursts,
i.e., no dependency on $L_{IR}$.  In contrast, \citet{lu03} found a
steady decrease of the PAH strength with increasing IR-activity.
Figure~\ref{figpahews} shows the equivalent widths of both the
$6.2\mu$m and $7.7\mu$m PAH as a function of the total infrared
luminosity for our starburst sample.  Within the uncertainties,
indicated by the scatter between the $6.2\mu$m and $7.7\mu$m PAHs for
the same object, the PAH equivalent widths remain constant over a
factor of 50 in total luminosity.  This finding is in good agreement
with \citet{pee04} (and references therein) who found that the
fraction of the total PAH flux emitted in the $6.2\mu$m PAH band
varies only slightly with an average of $28\%\pm 4\%$.  In other
words, the PAH flux and the underlying warm dust continuum scale
proportionally, and the PAHs and dust must be well mixed, at least on
large scales, to show these correlations.  This rules out a scenario of
a luminous, dusty nucleus surrounded by a large PDR with little
extinction, in favor of smaller, clumpier structures.

\subsection{PAH Luminosity and Star Formation Rate}
\label{pahsfr}

\citet{ken98} has shown that the $8-1000\mu$m infrared luminosity
$L_{IR}$ of starbursts is a good measure of the star formation rate
(SFR) given by 
\begin{displaymath}
\mbox{SFR}[M_{\odot}\mbox{yr}^{-1}] = 4.5\times
10^{-44} L_{IR} [\mbox{erg\ s}^{-1}].
\end{displaymath}
The SFR determines the number of young, massive stars, which provide
the (far-)UV photons to excite both PAH molecules and dust grains.  If
both species get excited by the same photons PAHs could potentially
be used as quantitative tracers of the SFR (see
section~\ref{secbrandllumi} for the dust luminosity).

Generally, PAHs are considered the most efficient species for
photoelectric heating \citep{bak94} in the PDRs. Molecular gas in
these boundary layers, surrounding the \2 regions, is exposed to
far-UV radiation ($6-13.6$\,eV), which strongly influences its
chemical and thermal structure \citep{tie85}.  PAHs are stochastically
heated by these UV photons, predominantly originating from massive
stars, and hence expected to be good tracers of star formation.  On
large angular scales -- similar to our case -- \citet{for04} found
that the $5-8.5\mu$m PAH emission constitutes an excellent indicator
of the star formation rate in circumnuclear regions and starbursts as
quantified by the Lyman continum flux, i.e. in regions where the
energy output is dominated by massive star formation.  However, it has
been known for a long time that PAHS can also be excited by visible
photons (e.g., \citet{uch98}), and that PAHs can trace other sources
besides massive young stars, such as planetary nebulae and reflection
nebulae.  If the observed PAH flux is integrated over the whole galaxy
it may predominantly trace B stars, which dominate the Galactic
stellar energy budget, rather than very recent massive star formation
\citep{pee04}.  This is in agreement with higher angular resolution
observations of the $3.3\mu$m PAH feature at the {\sl VLT} by
\citet{tac05}, who found a decrease in the PAH/continuum ratio at the
sites of the most recent star formation.

\begin{figure}[ht]
\epsscale{1.2}
\plotone{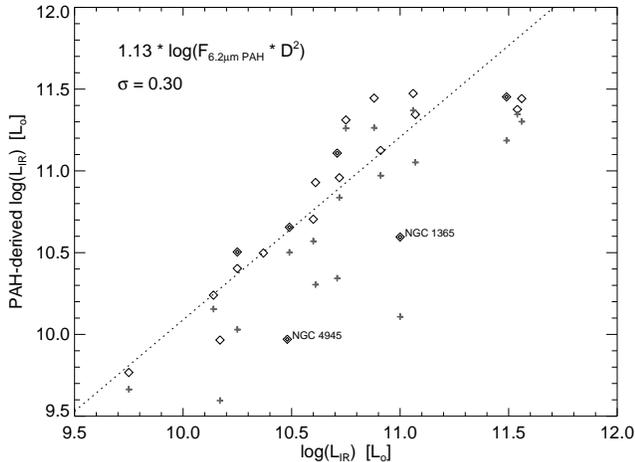}
\caption{The estimated infrared luminosity $L_{IR}^{PAH}$ based on the
         aperture corrected $6.2\mu$m PAH flux versus the total
         infrared luminosity as a measure of SFR.  The gray plus signs
         refer to the location of the same data points if no aperture
         correction is applied.  The dashed line represents a linear
         fit to the data (in log-log space), and the fitted function
         is given in the upper left, together with the standard error
         $\sigma$.  The labeled outliers NGC~1365, NGC~4945 were
         excluded from the fit.  The filled symbols correspond to
         starbursts with a weak AGN component. \label{figpahsfr}}
\end{figure}

In section \ref{secpahmixture} we have seen that the PAH equivalent
width does not dependent on $L_{IR}$.  However, that does not
necessarily mean that PAHs aren't good quantitative tracers of star
formation if the PAH feature and the underlying continuum scale
proportionally.  In Fig.~\ref{figpahsfr} we compare the flux in the
$6.2\mu$m PAH feature against the total infrared luminosity $L_{IR}$
from {\sl IRAS}.  To search for a physically meaningful correlation
one needs to take the distance of the object into account as well as
the fact that the narrow {\sl IRS} slit misses some of the total flux.
Hence we multiply the measured PAH fluxes with the square of the
distances and divide by the fractional flux factor FF
(Table~\ref{tabobsprop}).  A remarkably good fit can been achieved
with
\begin{displaymath}
\log\left(L_{IR}^{PAH}\right) = 1.13\times \log\left( F_{6.2\mu m PAH}
                                D^2 \right)
\end{displaymath}
where $F_{6.2\mu m PAH}$ is the $6.2\mu$m PAH flux in units of
$10^{-19}\mbox{W\,cm}^{-2}$, $D$ the distance in kiloparsecs, and
$L_{IR}$ in units of $L_{\odot}$.  The standard error is 0.3 in
$\log(L_{IR})$.  Using the above equation, the total infrared
luminosity of a starburst galaxy can be derived from the strength of a
single PAH emission feature (here: the $6.2\mu$m PAH) to within a
factor of two.  This correlation is less tight than the estimate from
the $F_{15\mu m}$ and $F_{30\mu m}$ continuum fluxes
(section~\ref{secbrandllumi}), supporting the finding by \citet{pee04}
that PAHs may not (only) trace recent massive star formation.

Combining the information from Figs.~\ref{figpahews} and
\ref{figpahsfr} suggests that the continuum and the PAH emission are, to
first order, proportional.  Similarly, \citet{pee04} have found that
$L_{FIR}$ is proportional to the PAH luminosity $L_{6.2\mu m}$.  If
our spectra were ``contaminated'' by a significant amount of continuum
emission from an AGN or an underlying, older galactic population
unrelated to the starburst, the equivalent width would vary with
distance (equivalent slit width), which is not observed.  Hence, we
conclude that both PAH and continuum emission originate predominantly
from the starburst.


\subsection{PAH Strength and Radiation Field}
\label{secpahvari}

It is well known that the equivalent width of PAHs is much reduced in
AGN-dominated environments (e.g., \citet{stu00}, \citet{gen00},
\citet{wee05}).  \citet{geb89} have discussed the susceptibility of
PAHs to destruction by far UV fields, and {\sl ISO} observations (e.g.,
\citet{ces96}, \citet{tra98}) have shown that more intense far-UV
radiation fields may lead to gradual destruction of PAHs around
stellar sources.  Recently, \citet{ywu06} studied a sample of
low-metallicity blue compact dwarf galaxies from $1/50 Z_{\odot}$ to
$1/1.5 Z_{\odot}$ with the {\sl IRS}.  They find a strong
anti-correlation between strength (equivalent width) of the PAH
features and the product of the [Ne\,III]/[Ne\,II] ratio (as a
hardness measure of the radiation field) and the UV luminosity density
devided by the metallicity.  A similar trend has been reported by
\citet{mad06} for a small sample of nearby dwarf galaxies.
Unfortunately, lower metallicity and harder radiation fields seem to
go hand-in-hand in these dwarf galaxies, and one cannot unambiguously
distinguish between possibly suppressed PAH formation in low
metallicity environments and PAH destruction in harder UV fields.
Recently, \citet{bei06} have investigated the strength of the
$11.3\mu$m PAH feature in the starburst in NGC~5253 for different
radial distances, and found that the equivalent width of the PAH
feature is inversely proportional to the intensity of the radiation
field, suggesting photo-destruction of the aromatic carriers in
harsher environments.

Observations of Galactic sources (e.g., \citet{ver96}, \citet{ver02})
have also shown that the relative strengths of individual PAH features
can depend on the degree of ionization of the molecule: C-C stretching
modes at $6.2\mu$m and $7.7\mu$m are stronger in ionized PAHs, while
the C-H in-plane bending mode at $8.6\mu$m and the C-H out-of-plane
bending mode at $11.3\mu$m are stronger by more than a factor of two
in neutral PAHs.  Comparing the {\sl ISOCAM} spectra of M\,82,
NGC~253, and NGC~1808, \citet{for03} found that, while the $5-11\mu$m
spectrum is nearly invariant, the relative PAH intensities exhibit
significant variations of $20\% - 100\%$, which they attributed to the
PAH size distribution, ionization, dehydrogenation, or the incident
radiation field.

\begin{figure}[ht]
\epsscale{1.2}
\plotone{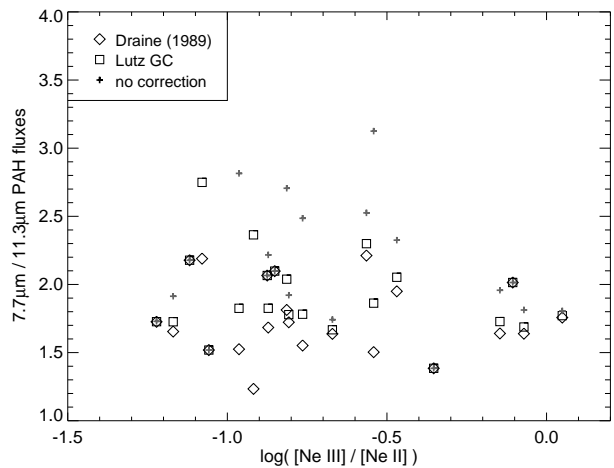}
\caption{The ratio of the $7.7\mu$m\,/\,$11.3\mu$m PAH fluxes versus
         the [Ne\,III]/[Ne\,II] line ratio (from \citet{dev06}).  The
         PAH fluxes have been corrected for extinction using the laws
         from \citet{dra89} (diamonds) and \citet{lut99} (squares).
         The gray pluses indicate the locations of the uncorrected
         values for reference.  The filled diamonds correspond to
         starbursts with weak AGN components.\label{figpahrationeon}}
\end{figure}

In Fig.~\ref{figpahrationeon} we compare the ratio between the
$11.3\mu$m bending mode and the $7.7\mu$m stretching mode to the
hardness of the radiation field, as indicated by the
[Ne\,III]/[Ne\,II] ratio from \citet{dev06}, who provide a detailed
analysis of the fine structure lines in our starburst sample.  We
correct both PAH fluxes for extinction (section~\ref{secpahmixture})
using the extinction laws by \citet{dra89} and \citet{lut99} and the
values from Table~\ref{tabextlaws}.  Within the systematic
uncertainties, represented by the scatter of the data points, we find
no significant variation of the $7.7\mu$m\,/\,$11.3\mu$m PAH ratio
over more than an order of magnitude in [Ne\,III]/[Ne\,II] fluxes.
While we cannot exclude variations on scales of individual \2 regions,
our spatially averaged spectra of starburst nuclei do not reveal
significant variations between the main PAH features.  We have also
looked at the weaker PAH features from Table~\ref{tabpahs}, but the
scatter increases with lower signal-to-noise and more uncertain
baseline definition, and does not reveal an obvious trend.  However,
more studies, in particular of the PAH complex around $17\mu$m are
planned.

\begin{figure}[ht]
\epsscale{1.2}
\plotone{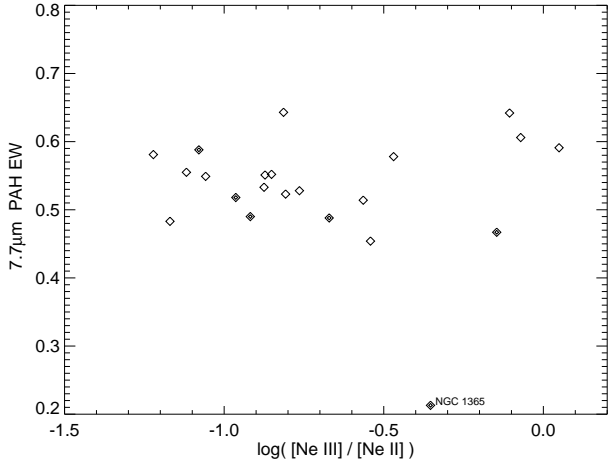}
\caption{The equivalent width of the $7.7\mu$m PAH feature versus the
         [Ne\,III]/[Ne\,II] line ratio (from \citet{dev06}) as an
         indicator of the hardness of the radiation field.  The filled
         diamonds correspond to starbursts with weak AGN
         components. NGC~1365, which has a significant AGN component,
         is a clear outlier.
\label{figpahewsneon}}
\end{figure}

Our starburst sample spans a wide range in radiation field hardness.
Fig.~\ref{figpahewsneon} shows the equivalent width of the $7.7\mu$m
PAH feature versus the fine structure line ratio [Ne\,III]/[Ne\,II],
which has been taken from \citet{dev06}.  Within the uncertainties the
equivalent width of the $7.7\mu$m PAH feature remains constant over
more than an order of magnitude in [Ne\,III]/[Ne\,II].  We conclude
that, {\sl on large scales} of starburst nuclei, which typically
contain numerous \2 regions, the PAH~/~continuum ratio does not
significantly depend on the average radiation field hardness.

\section{Summary} 
\label{secsummary}

We presented and discussed the $5 - 38\mu$m mid-infrared spectra of a
large sample of 22 starburst galaxies taken with the Infrared
Spectrograph {\sl IRS} on board the Spitzer Space Telescope.  The high
signal-to-noise spectra contain numerous important diagnostics such as
PAH emission features, silicate bands at $9.8\mu$m and $18\mu$m, and
the shape of the spectral continuum.  The {\sl IRS} spectral
resolution of $R\approx 65 - 130$ is perfectly matched to study these
features.  From our sample we constructed an average starburst
spectrum, which can be used as a starburst template.

The availability of continuous mid-infrared spectra of numerous
objects within one class over a wide wavelength range enables various
important studies.  Remarkably, the spectra show a vast range in
starburst SEDs.  We found a trend that more dust extincted starburst
systems have a steeper spectral continuum slope longward of $15\mu$m.
The slope can also be used to discriminate between starburst and AGN
powered sources, with a transition at $0.17\le F_{15\mu m}/F_{30\mu m}
\le 0.21$.  The monochromatic continuum fluxes, which represent a more
accurate estimate of the ``true'' continuum than broad band filters,
provide a remarkably accurate estimate of the total infrared
luminosity via $L_{IR}^{est} = D^2\cdot (4.27\cdot F_{15\mu m} +
11\cdot F_{30\mu m})$ (after correcting for slit losses for nearby,
extended systems).

Our starburst spectra cover a wide range of silicate absorption
depths, from essentially no absorption to heavily obscured systems
with an optical depth of $\tau_{9.8\mu m}\sim 5$.  We present the
discovery of crystalline silicates in NGC~4945, which shows many
similarities with heavily extincted ULIRGs.  However, unlike the
latter, the starbursts in our sample show no signs of water ices or
hydrocarbons, suggesting a small amount of self-shielding.

The observed spectra show significant variations in the relative
strengths of the individual PAH features at $6.2\mu$m, $7.7\mu$m, and
$11.3\mu$m.  However, these variations may be entirely due to
extinction and do not necessarily indicate intrinsic variations of the
PAH spectrum.  We find that the PAH equivalent width is independent of
the total luminosity $L_{IR}$, probably because the PAH strength and
the underlying continuum scale proportionally within a ``pure''
starburst.  The luminosity of an individual PAH feature, however, scales
with $L_{IR}$.  In particular the $6.2\mu$m feature can be used to
approximate the total infrared luminosity of the starburst (although
less accurately than from the $15\mu$m and $30\mu$m continuum fluxes).

We investigated possible variations of the PAH spectrum as expected,
e.g., from varying degrees of PAH ionization.  The
$7.7\mu$m\,/\,$11.3\mu$m PAH ratios show no significant systematic
variation with the hardness of the radiation field.  Although our
sample covers about a factor of ten difference in radiation field
hardness (as indicated by the [Ne\,III]\,/\,[Ne\,II] ratio) we found no
systematic correlation with the PAH equivalent width.  Furthermore, we
found no systematic differences between ``pure'' starbursts and
galaxies with a weak, non-dominant AGN component for most of their
spectral properties (except for NGC~1365 which shows very weak PAH
emission).

We emphasize that these results are based on spatially integrated
diagnostics over an entire starburst region.  Local variations of age,
IMF, density or geometry on the scales of individual \2 regions or
super star clusters may just average out.  However, it is important to
note that, because of this ``averaging out effect'' in unresolved
sources, starburst nuclei with significantly different global
properties may appear as rather similar members of one class of objects.


\acknowledgments

This work is based on observations made with the {\em Spitzer} Space
Telescope, which is operated by the Jet Propulsion Laboratory,
California Institute of Technology under NASA contract 1407. Support
for this work was provided by NASA through Contract Number 1257184
issued by JPL/Caltech.



\end{document}